%% file: maxima_v9.2.tex
\documentclass[a4paper,10pt]{article}
\pdfoutput=1
\usepackage[margin=1.2in]{geometry}
\usepackage[utf8]{inputenc}
\usepackage{acronym}
\usepackage{amsmath}
\usepackage{amsfonts}
\usepackage{amssymb}
\usepackage{authblk}
\usepackage{mathrsfs}
\usepackage{graphicx}
\usepackage{hyperref}

\newcommand{\Tc}{{T_\mathrm{c}}}
\newcommand{\bc}{{\beta_\mathrm{c}}}
\newcommand{\bi}{{\beta_\mathrm{i}}}

\newcommand{\md}{\mathrm{d}}
\newcommand{\Eth}{{E_\mathrm{th}}}
\newcommand{\MAX}{\mathrm{max}}
\newcommand{\MIN}{\mathrm{min}}
\newcommand{\erfc}{\mathrm{erfc}}

\newcommand{\emax}{E_\MAX}
\newcommand{\emaxt}{E_\MAX(t)}
\newcommand{\emin}{E_\MIN}
\newcommand{\emint}{E_\MIN(t)}

\newcommand{\rhogauss}{{\rho_\mathrm{gauss}}}

\graphicspath{{./FIGURE/}{./}{../plots/}}

\acrodef{rem}[REM]{Random Energy Model}
\acrodef{tm}[TM]{Trap Model}
\acrodef{mf}[MF]{mean field}
\acrodef{mc}[MC]{Monte Carlo}
\acrodef{rsb}[RSB]{Replica Symmetry Breaking}
\acrodef{iid}[i.i.d.]{independent identically distributed}

\title{Maximum-energy records in glassy landscapes}
\author[1]{Ivailo Hartarsky\thanks{\textsf{ivailo.hartarsky@ens.fr}}}
\author[2]{Marco Baity-Jesi}
\author[3]{Riccardo Ravasio}
\author[4]{Alain Billoire}
\author[5]{Giulio Biroli}
\affil[1]{\small D\'epartement Mathématiques et Applications, \'Ecole Normale Sup\'erieure, CNRS, PSL Research University, 75005 Paris, France}
\affil[2]{\small Chemistry Department, Columbia University, 3000 Broadway, 10027 New York (NY), USA}
\affil[3]{\small Institute of Physics, \'Ecole Polytechnique F\'ed\'erale de Lausanne, CH-1015 Lausanne, Switzerland}
\affil[4]{\small Institut de Physique Th\'eorique, Universit\'e Paris Saclay, CNRS, CEA, F-91191, Gif-sur-Yvette, France}
\affil[5]{\small Laboratoire de Physique Statistique, École Normale Supérieure, CNRS, PSL Research University, 75005 Paris, France}
\begin{document}

\maketitle

\begin{abstract}
We study the evolution of the maximum energy $\emaxt$ reached between time $0$ and time $t$ in the dynamics of simple models with glassy energy landscapes, in instant quenches from infinite temperature to a target temperature $T$.
Through a detailed description of the activated dynamics, we are able to describe the evolution of $\emaxt$ from short times, through the aging regime, until after equilibrium is reached, thus providing a detailed description of the long-time dynamics. Finally, we compare our findings with numerical simulations of the $p$-spin glass and show how the maximum energy record can be used to identify the threshold energy in this model.
\end{abstract}


\tableofcontents

\section{Introduction}

Glass formers incur a sensational dynamical slowdown as the temperature is decreased~\cite{debenedetti:01}. At high temperatures, this slowdown is well explained by \ac{mf} theory~\cite{cavagna:09,maimbourg:16}, which predicts a breaking of ergodicity at a temperature $T_\mathrm{d}$: at $T<T_\mathrm{d}$ the typical configurations are confined in local minima of the energy landscape. Since, in \ac{mf}, exiting these local minima requires overcoming barriers of height diverging with the system size $N$, in the thermodynamic limit the system remains stuck in a portion of phase space.

Because in non-\ac{mf} systems the energy barriers are not diverging, real low-dimensional glasses are able to escape local minima of the energy when $T<T_\mathrm{d}$, and ergodicity is restored. Nevertheless, the dynamics below $T_\mathrm{d}$ is extremely slow, since overcoming a barrier of height $\Delta$ requires an exponentially large time $\tau$, given by the Arrhenius law, $\tau\propto\exp(\Delta/k_\mathrm{B}T)$. 
This extremely slow low-temperature dynamical regime, consisting in jumps over energy barriers, is often referred to as (thermal) \emph{activation} or activated dynamics.

Activated dynamics can be studied in \ac{mf} models by considering exponentially large times in $N$~\cite{crisanti:00,baityjesi:18,stariolo:19}, which restricts numerical simulations to very small system sizes. From the analytical point of view, the situation is even harder, because the thermodynamic limit cannot be taken, since once $N$ is sent to infinity it becomes impossible to treat times of order $\exp(N)$. A consequence is that, to our knowledge, there are very few exact analytical calculations for the activated regime in canonical glass models.

A successful alternative analytical framework for the description of activated dynamics is the \ac{tm}~\cite{dyre:87,bouchaud:92}, a toy model that describes the simplest case of activation, in the absence of any other dynamical mechanism, where local minima (traps) are exited by reaching a threshold energy $\Eth$. A detailed non-trivial analysis has been made for the \ac{tm}~\cite{bouchaud:92,bouchaud:95,dyre:95}, and a series of works culminated recently has shown that for large enough system sizes the \ac{tm} correctly describes the activated long-time dynamics of a paradigmatic glass system, the \ac{rem}~\cite{benarous:02,benarous:08,gayrard:16,gayrard:18,baityjesi:18}.

In this paper we go beyond these works in two main ways. Firstly, we thoroughly characterise the dynamics of the \ac{rem} at \emph{all} time scales. Secondly, we unveil an interesting behaviour of the maximum reached energy $\emaxt$, which can be related to static quantities such as the threshold energy $\Eth$. Our findings are then tested by numerical simulations in the $p$-spin model, which provides a more complex case of glassy dynamics.

Our approach is based on record statistics, which addresses questions on the extreme events arising in series of random variables~\cite{gumbel:58,schehr:13}. The simplest case of tracking the maximum in a series of
independent random variables is well understood, but 
as soon as some dependence is added to the problem, the results become less trivial, and rigorous results
are only known for a limited number of problems, such as random walks \cite{schehr:13} (see \cite{godreche:17} for a review) and
eigenvalue statistics \cite{johnstone:08,tracy:94,tracy:96} (see \cite{majumdar:14} for a review).

We focus on three paradigmatic models of mean-field spin glasses, the \acf{tm}, the \acf{rem}, and the $p$-spin
model, with particular emphasis on the \ac{rem}. We study the evolution of the maximum energy, $\emaxt$, reached by time $t$ after an instant quench from an infinite to a small temperature $T$.

The \ac{tm} consists of a random walk in a fully-connected graph. The graph is equipped with a random potential and the transition rates do not depend on the destination energy. For this model we find that $\emaxt$ has a simple behaviour at all times.

In the \ac{rem}, the random walk is on a more sparse graph and the transition rates depend on the difference between the starting and destination energy. These features lead to a rich non-trivial behaviour of $\emaxt$, in which we are able to identify several different regimes for different lapses of time from the quench (see Sec.~\ref{subsec:summary} for a summary and Tab.~\ref{tab:summ} for more detailed results). Following $\emaxt$ opens a window on exponentially large time scales which are usually hard to study.

Finally, we make a digression to the $p$-spin model, whose landscape is more complex than the one of the \ac{rem}. This is due to an explicit correlation among the energy levels, which are no longer \ac{iid} random variables. We show numerically that even in this case the maximum energy reached grows
logarithmically. We also argue that the maximal energy $\emax(t/2,t)$, reached in the time interval $[t/2,t]$, can be used to identify the threshold energy $\Eth$ (i.e. the highest energy at which local minima are found) in a finite size system. We further show numerically that this estimate converges smoothly to the analytical prediction in the thermodynamic limit.

In section \ref{sec:models} we define the models we study. In section \ref{sec:trap} we describe the evolution of the maximum energy reached in the \ac{tm}, whereas in section \ref{sec:rem} it is analysed in the \ac{rem}. In section \ref{sec:pspin} 
we discuss the $p$-spin model and we show how the maximum energy reached can be used to locate the threshold energy. Finally, we give some concluding remarks in section \ref{sec:conclusions}.
In appendix~\ref{app:I} we remind the reader some standard extreme-value statistics results, and in appendix~\ref{app:finitecorr} we take into account a number of subleading corrections which allow for a quantitative comparison of our results with simulations on very small systems.

\section{Models}\label{sec:models}
In this section we describe the three models treated in this paper.

\subsection{Trap model}
In the \ac{tm}~\cite{dyre:87,bouchaud:92, monthus:96}, $M$ states lie on a fully connected graph, and their energies are drawn independently from
an exponential distribution
\begin{equation}\label{eq:rhoexp}
 \rho_\mathrm{exp}(E) = \alpha e^{\alpha E} \Theta(-E)\,,
\end{equation}
where $\Theta(\cdot)$ is the Heaviside step function and $\alpha>0$ is a parameter that will be set to 1. 

The transition rate, $q_{i,j}$, from a state $i$ to a state $j$ does not depend on the target state $j$:
\begin{equation}\label{eq:rateTrap}
q_{i,j}=\frac{1}{M}e^{\beta E(i)}\,,
\end{equation}
where $\beta=1/T$ is the inverse temperature and $E(i)$ is the energy of state $i$.
This represents a system in which leaving a state (or \emph{trap}) requires reaching a threshold energy $\Eth=0$. Once the threshold is reached, the whole space of states becomes accessible. This model has weak ergodicity breaking at temperatures $T\leq\Tc=\frac{1}{\alpha}$ \cite{bouchaud:92} (see also paragraph \ref{sec:emaxTM}).

Summarizing, in the \ac{tm} the energies of different states are independent,
the space of states is fully connected, 
and the dynamics depends only on the state issuing the transition and not on the target state,
implying that there is a constant threshold energy $\Eth=0$ separating the configurations.

\paragraph{Gaussian Trap model}
The \ac{tm} can be defined analogously for a Gaussian energy distribution
\begin{equation}\label{eq:rhogauss}
 \rhogauss(E) = \frac{1}{\sqrt{2\pi N}}e^{-\frac{E^2}{2N}}\,,
\end{equation}
setting $M=2^N$.
In this case, also positive energies can be reached, but since $\Eth=0$ the dynamics leaves them quickly with a rate given by equation \eqref{eq:rateTrap}.

\subsection{Random Energy Model}
The \ac{rem} describes a system of $N$ binary spins \cite{derrida:80,derrida:81}. As a consequence, the total number of states is $M=2^N$, and from
each one it is possible to reach $N$ new states by flipping a single spin. The energy change induced by flipping any of the spins is assumed to be so drastic that the energy levels are independent from each other. The energy $E$ of a state is drawn from
the probability distribution $\rhogauss$ (see Eq.~\eqref{eq:rhogauss}).

The statics of the \ac{rem} can be solved exactly. There is a disordered phase for temperature $T>\Tc=\frac{1}{\sqrt{2\ln(2)}}$,
and a glassy phase for $T<\Tc$~\cite{derrida:80,derrida:81}.~\footnote{
The critical temperature $\Tc$ is different from the one given in Refs.~\cite{derrida:80,derrida:81}, because we define the model with variance $\mathrm{Var}(E)=N$, instead
of $\mathrm{Var}(E)=N/2$.}
Concerning the dynamics of the \ac{rem}, we will focus on \ac{mc} Metropolis dynamics.
More precisely, we consider that spins flip one at a time and we set the transition rate from state $i$ to state $j$, differing by a single spin, to be
\begin{equation}\label{eq:mc}
q_{i,j}=\frac{1}{N}\min\left(1,e^{\beta (E(i)-E(j))}\right)\,. 
\end{equation}

In a nutshell, the \ac{rem} can be seen as a \ac{tm} with (i) a hypercubic space of configurations (instead of fully connected), and (ii) with more physical dynamics to move in it. 
In the limit of very long times and large system sizes, the dynamics of the \ac{rem} is qualitatively equivalent to the one of the \ac{tm}, with a threshold energy $\Eth=-\sqrt{2N\ln(N)}$ \cite{baityjesi:18, gayrard:18}.

\subsection[The \texorpdfstring{$p-$}-spin model]{The \texorpdfstring{$\boldsymbol{p}$}{p}-spin model} \label{sec:pspin-def}
The $p$-spin model represents a system of $N$ binary spins $\sigma_i=\pm1$ \cite{derrida:80}. At variance with the \ac{tm} and the \ac{rem}, where the energies are i.i.d., 
the energy depends on the microscopic configuration of the system through the Hamiltonian
\begin{equation}
\label{eq:pspin-H}
H = -\sum_{i_1<\ldots<i_p} \ J_{i_1,\ldots,i_p}\sigma_{i_1}\ldots\sigma_{i_p} \ ,
\end{equation}
in which the interactions couple all the possible groups of $p$ different spins. 
The bonds $J_{i_1,\ldots,i_p}$, associated to a $p$-tuple of spins, are independent and usually follow a Gaussian or bimodal distribution. We choose the latter, with $J_{i_1,\ldots,i_p} = \pm \sqrt{p!/N^{p-1}}$ 
with probability one half.
An alternative way to define the previous Hamiltonian is as a Gaussian random field defined on the $N$-dimensional hypercube (in this case the couplings are Gaussian random variables). Its mean is zero, whereas its covariance in the large $N$ limit is {$\overline{H(\{\sigma_i\})H(\{\sigma_i'\})}\sim N q^p$} where $q=\sum_i \sigma_i\sigma_i'/N$.

When $p$ is finite, there is a general agreement that all models with finite $p\geq3$ fall in the same universality class \cite{billoire:05}.
Instead, if $p$ is diverging, the way limits are taken becomes important. On the one hand, if one studies the short-time limit, sending first $N\to\infty$, and only later $p,t\to\infty$, the $p$-spin reduces to the \ac{rem}~\cite{derrida:80} (which corresponds to naively saying that in the large $N$ limit the covariance tends to $\delta_{\{\sigma_i\}\{\sigma_i'\}}$).
On the other hand, when studying the long-time regime of activated dynamics, one needs to study large but finite $N$ at every time $t$ (recall that by definition $p\leq N$), so the REM cannot be recovered~\cite{baityjesi:18c}.

We will focus on $p=3$, in the limit of large but finite $N$, where activated processes become possible \cite{crisanti:00}.
As for the dynamics, we use the single-spin \ac{mc} Metropolis dynamics, described in Eq.~\eqref{eq:mc}.

\section{Energy maxima in the Trap Model}\label{sec:trap}

\subsection{Qualitative description of the dynamics in the TM}
\label{sec:emaxTM}
We consider a typical non-equilibrium protocol: an instant quench from infinite temperature to a target temperature $T$.
In the TM, the system is always stuck in a trap, where it will remain for a time $\tau\propto e^{-\beta E}$ (Eq.~\eqref{eq:rateTrap}), before emerging to the threshold to transition to the next trap.
We call $\tau$ the trapping time.

One can easily see that the distribution of trapping times is {heavy-tailed with density}
\begin{equation}
\psi(\tau)\propto\tau^{-(\alpha T+1)}\,,
\end{equation}
which identifies a critical temperature at $\Tc\equiv\frac{1}{\alpha}$~\cite{bouchaud:92}. 
When $T<\Tc\equiv\frac{1}{\alpha}$ the average trapping time is infinite and the total waiting time is of the order of the maximal waiting time in a single state~\cite{bouchaud:92}. 
Consequently, the process waits in the deepest trap it has seen for about as much time as the system has spent in the heat bath, implying
\begin{equation}\label{eq:et}
E(t)\approx\emint\sim-T\ln t\,.
\end{equation}
Eventually, the system is able to reach $\Eth$ and fall back into shallower traps, until it finds an even deeper trap than the one before, where it spends a time longer than previously elapsed time. 
As a consequence, as time passes, the energy decreases gradually, and the dynamics becomes increasingly slower.
This dependence on how long the system has 
been in the bath is called \emph{aging} and results in an increasing time correlation of the system.
In finite systems aging goes on until equilibrium is reached. 

\subsection{Energy Maxima in the Exponential Trap Model}
For our consideration of the maximum energy, $\emaxt$, reached after a time $t$ we will restrict ourselves to the low-temperature \ac{tm}, where aging occurs. In other words, we assume that $T<\Tc$, so that \eqref{eq:et} holds. Then $\emaxt$ can be calculated through extremum statistics. 
Recalling that each new trap is chosen independently, the number of traps, $n_t$, that the system will have visited satisfies
\begin{equation}
\label{eq:tm_positive}
n_t \int_{-\infty}^{{E_\MIN(t)}} \rho_\mathrm{exp}(E) dE \sim 1\,,
\end{equation}
where $E_\MIN(t)$ is the minimum energy reached until time $t$. 
Therefore, $n_t\sim e^{-\alpha E_\MIN(t)}\sim e^{-\alpha E(t)}$, since in the \ac{tm} the current energy is almost always the lowest energy reached (i.e. $E(t)\approx E_\MIN(t)$).

The number of visited traps needs to satisfy an analogous relation for the highest energy reached
\begin{equation}
n_t \int_{\emaxt}^{0} \rho_\mathrm{exp}(E) dE \sim 1\,,
\end{equation}
that yields $n_t{\sim -1/\emaxt}$.

Equating the two expressions for $n_t$, and using \eqref{eq:et}, one obtains the behaviour
\begin{equation}
 \emaxt \sim \ln\left(1-t^{-\frac{T}{\Tc}}\right)\sim -t^{-\frac{T}{\Tc}}\,,
\end{equation}
so in the exponential \ac{tm} the largest energy reached approaches its extremal value $E=0$ as a power law.
\footnote{{The high-temperature case, $T>\Tc$, can be treated similarly, yielding $\emaxt\sim -1/t$.}}

Note that with respect to all the models defined below, the exponential \ac{tm} is special (and a bit pathological) since 
it has a cut-off in the energy distribution at the most numerous energies, i.e. it does not have any rare high energy. 
In consequence, its behaviour is atypical with respect to all other cases considered below.

\subsection{Gaussian Trap Model}
In the Gaussian \ac{tm}, since $\rho(E)$ is symmetric, the 
maximum and minimum records $\emaxt$ and $\emint$ follow the same law, with opposite signs.
This is because, from any configuration $i$, all the configurations are reached with equal probability,
and what changes is only the amount of time the system remains in them.
Therefore, a configuration with $E<\emint$ is reached with the same probability as a configuration with energy $E>-\emint$.
Since, by Eq.~\eqref{eq:et}, in the \ac{tm} $E(t)\sim\emint\sim-T\ln t$, we can deduce that
\begin{equation}
\label{eq:gtm:aging}
\emaxt\sim T\ln t.\end{equation}
{Eq.~\eqref{eq:et} still holds for the Gaussian \ac{tm} in its aging regime, $t<\exp(\beta\bi N)$, where $\bi=\min(\beta,\bc)$ and $\bc=\sqrt{2\ln 2}$. If $\beta<\bc$ (i.e. in the high-temperature phase), a supplementary `post-equilibration' regime develops for $\exp(\beta^2 N)<t<\exp(N(\bc^2+\beta^2)/2)$, governed by
\begin{equation}
\label{eq:gtm:equilibrium}
\emaxt=\sqrt{2N(\ln t-\beta^2N/2)}.
\end{equation}
In both cases, the evolution of the maximum energy eventually ends when $\emax$ reaches the maximum possible value
$N\sqrt{2 \ln 2}$. 
As we shall discuss a similar argument for the \ac{rem} in Sec.\ref{subsec:verylongscales} and since there is no difference with the minimal energy, we omit the details here. However, let us note that depending on the temperature there are two different behaviours of $\emaxt/N$ as a function of $\ln t/N$. For $\beta>\bc$ there is a single regime, linear in $\ln t$~\eqref{eq:gtm:aging}, while otherwise there are two different ones -- a linear law~\eqref{eq:gtm:aging} followed by a square-root dependency~\eqref{eq:gtm:equilibrium}.}

\section{Energy maxima in the REM}
\label{sec:rem}

We now turn to the Metropolis dynamics on the \ac{rem}. As in the previous section we examine quenches from infinite temperature to a target temperature $T$. Throughout this section we consider the thermodynamics large-$N$ limit and systematically discard subleading terms. Higher-order corrections for finite systems are discussed in App.\ref{app:finitecorr}.

\paragraph{Short summary of the maximum-energy records in the \ac{rem}.}
\label{subsec:summary}
We are able identify and study a large number of different regimes in the dynamics, which are summarised in quantitative terms in Tab.~\ref{tab:summ} at the end of Sec.~\ref{sec:rem}.

The energy of the initial state is typically of order $\sqrt{N}$ (Sec.~\ref{subsec:init}). In the initial stages of the evolution, the system quickly falls into a local minimum (Secs.~\ref{subsec:drift} and~\ref{subsec:firsttrap}). Since it is not necessary to reach energies of order $\sqrt{N}$ to escape the local minima, $\emaxt$ stays unchanged and only starts growing once a time of order $t_3=e^{\sqrt{2\beta^3 N\sqrt{2N\ln N}}}$ has passed (Eq.~\eqref{eq:e1}, Sec.~\ref{subsec:reachingcommonenergies}).

At larger times, and, in particular, those exponentially large in $N$, $\emaxt$ incurs a further slow down, because not only new records are hardly accepted, but they are also harder to find (Sec.~\ref{subsec:aging}). Even after equilibrium has been reached, $\emaxt$ keeps growing (Sec.~\ref{subsec:verylongscales}), and saturates only when the global maximum of the system has been visited (Sec.~\ref{subsec:saturation}).

\subsection{Basic features of the landscape and the dynamics of the REM}
In the \ac{rem}, most energy levels are concentrated around zero at a typical distance of order $\sqrt{N}$. 
At variance with the \ac{tm}, the phase space has a structure, and each state has $N$ neighbours
whose energies,
as we show in Sec.~\ref{subsec:asymptoticbounds}, Eq.~\eqref{eq:Iasymp}, {typically}
lie in the interval
\begin{equation}
\label{eq:Ieqiv}
I\equiv\left[-\sqrt{2N\ln N},\sqrt{2N\ln N}\right]\,.
\end{equation}
The majority of states have at least one lower neighbour, and the dynamics spends little time there, since the Metropolis update rule [Eq.~\eqref{eq:mc}] privileges energy descent.
As a consequence, the states in which most of the time is spent are those without a lower neighbour, 
which have energy $\inf I=-\sqrt{2N\ln N}$ or lower. Minima with energy $E\sim-N$ can be seen as the equivalent of the \ac{tm} traps,
since $N\gg\sqrt{2N\ln N}$ (the separation between these minima and the threshold diverges with $N$). Since the energy of such states is much lower than the typical energies, to leading order the time spent in these low-energy minima configurations scales as $\tau\sim e^{-\beta E}$ [Eq.~\eqref{eq:mc}].

\subsection{Energy maximum evolution}
In this section, as we did for the \ac{tm}, we discuss the evolution of the typical maximum energy $\emaxt$ in the \ac{rem} in the limit of very large system sizes. The final results will be summarised in Table \ref{tab:summ}.

When dealing with exponentially large time scales,
a central assumption of our computation is that, each time the dynamics leaves a state, the system becomes independent from its past. In other words, the neighbours are drawn anew, so that returns to a recently visited configuration are not taken into account. This hypothesis is supported by an argument in Ref.~\cite{baityjesi:18}, stating that even though, in the actual dynamics, some configurations are revisited, this happens with a sub-exponential rate. In consequence, this should not affect exponential time scales.

\subsubsection{Initial condition}\label{subsec:init}
Before the quench the temperature is infinite, so $E_\MAX(0)$ is drawn from $\rhogauss(E)$. Therefore, its intensive value is zero in the limit of large system sizes: $E_\MAX(0)/N\sim1/\sqrt{N}\stackrel{N\to\infty}{\longrightarrow} 0$.

\subsubsection{Drift towards the first trap}
\label{subsec:drift}
\begin{figure}[!tbh]
\includegraphics[trim={0 13cm 0 0}, width=\columnwidth]{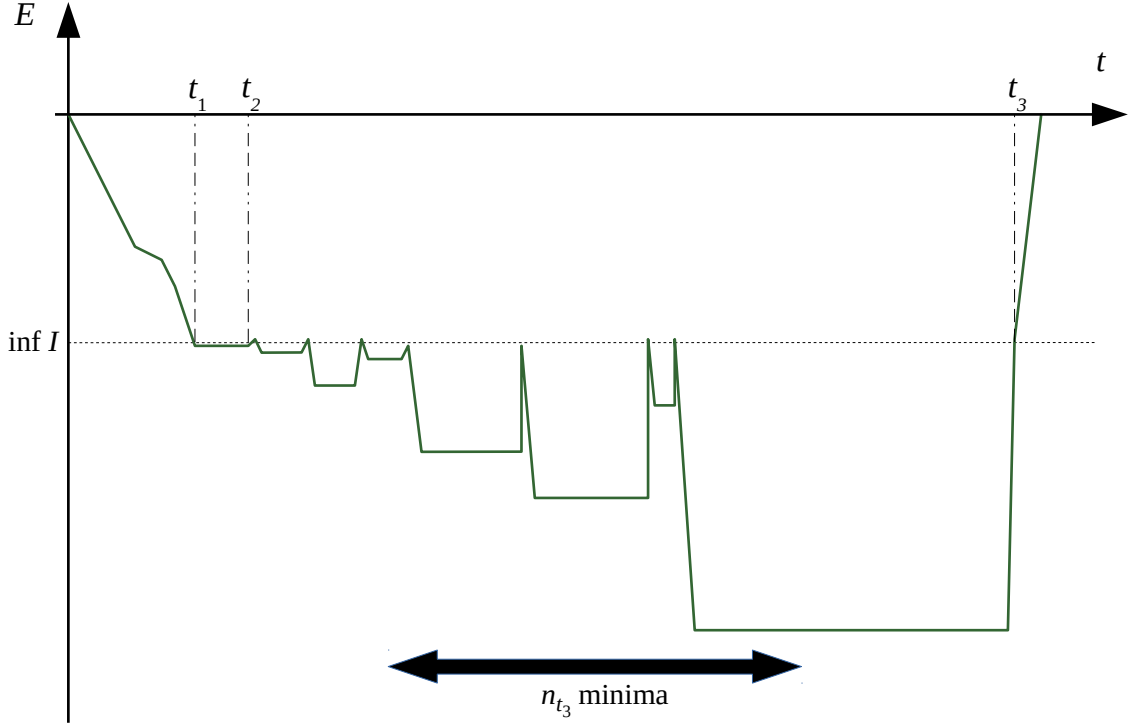}
\caption{A schematic representation of the dynamics of the \ac{rem} in the initial stages of the dynamics. In a time of order $t_1=N$ the system falls in the first local energy minimum, whose energy is typically $\inf I$ [Eq.~\eqref{eq:Ieqiv}]. It takes a time of order $t_2=\exp(\sqrt{N/\ln N})$ to leave that state. After that, the system will visit a large number, $n_{t_3}$, of minima before it can reach energies of order $\sqrt{N}$, after a time $t_3=e^{\beta E_1}$ [Eq.~\eqref{eq:e1}]. During all this time $\emaxt=E(0)$. The times and energies in the diagram are not drawn to scale.}
\label{fig:beginning}
\end{figure}

At the beginning of the evolution, the system quickly falls into a first local minimum of the energy. See Fig.~\ref{fig:beginning} (left part) for a schematic description of this and the following regimes.

In fact, with probability $1-\frac{1}{N+1}$, the initial state has at least one lower neighbour. 
The rate $q_{i,j}$ [Eq.~\eqref{eq:mc}] privileges energy descent, and on time-scales of order one the energy will typically immediately decrease since climbing up moves will be discarded (their rate is exponentially small in $\sqrt N$). The transition toward the lowest neighbour, which is at energy $\inf I$, has a rate $q_{i,j}\sim 1/N$, hence a local minimum will be reached in a time of order $t_1=N$ (a configuration at energy $\inf I$ is a local minimum with probability of order $1$).
The number of transitions required to reach such state is of order $\ln(N)$. \footnote{At each transition, the number of lower neighbours is typically divided by two, so it will take $\mathcal{O}(\log_2(N))$ steps to reach a local minimum.}

During this regime, before the first trap is reached, $E_{\max}$ maintains its initial value of order $\sqrt N$, because the energy is decreasing.

\subsubsection{Stay in the first trap}
\label{subsec:firsttrap}
Once the system reaches the first local minimum, which has energy $E\approx\inf I$ (that of a typical lowest neighbour), the energy difference with its neighbours is of order $\delta=\sqrt{\frac{N}{\ln N}}$ (see App.~\ref{subsec:lowestneighboursgap}). 
Therefore, the system remains in the first trap and does not move at all for a time of order $t_2=\exp\left(\beta \sqrt{\frac{N}{\ln N}}\right)$. Consequently, $\emaxt$ does not change either.

\subsubsection{Reaching common energies}
\label{subsec:reachingcommonenergies}
We now estimate the time scale $t_3$ required, once the dynamics left the first trap, for the energy to become of order $\sqrt{N}$. 
Such energy needs to be reached for a new record $\emaxt>E(0)\sim\sqrt{N}$ to be hit.

After $t_2$, the system starts visiting a series of local minima of the energy. When the dynamics leaves a local minimum, it does so by jumping to its lowest-energy neighbour, which, to leading order, also has energy $\inf I$. In order to reach energies of order $\sqrt{N}$, a large number of local minima needs to be visited.

The typical trapping time in such a minimum with energy $E$, is
\begin{equation}
\tau\sim Ne^{\beta\left(\inf I - E\right)}
\end{equation}
and the rate of jumping to a configuration of energy of order $\sim\sqrt{N}$ from the local minimum is
\begin{equation}
q_\mathrm{common}\sim e^{\beta\left(E-\mathcal{O}(\sqrt{N})\right)}\,,
\end{equation}
where $\mathcal{O}(\sqrt{N})$ indicates a quantity of order $\sqrt{N}$.
The probability of jumping in a time $\tau$ to a configuration of energy $\sim\sqrt{N}$ is then given by
\[
\tau q_\mathrm{com}\sim e^{\beta(\inf I-\mathcal{O}(\sqrt{N}))}\approx e^{\beta\inf I}\,.
\]

For such a transition to be likely, the number of visited traps $n_{t_3}$ needs to be of 
order $e^{-\beta\inf I}$, since any site usually has neighbours at energy $\sim\sqrt{N}$. The lowest of the $n_{t}$ traps determines $\emint$. Thus, $\emint$ can be calculated as the minimum among $Nn_{t}$ Gaussians (the factor $N$ stems from the fact that we are dealing with minima of $N$ Gaussians, not arbitrary Gaussians) with variance $N$. Therefore, through known extremum statistics results (see App.~\ref{subsec:asymptoticbounds}, Eq.~\eqref{eq:mn}), one obtains
\begin{equation}\label{eq:e1}
-E_{\min}(t_3)\sim\sqrt{2N\ln {\left(Nn_{t_3}\right)}}\approx \sqrt{2\beta N\sqrt{2N\ln N}}=:E_1.
\end{equation}
By using Eq.~\eqref{eq:et}, which is valid also in the \ac{rem} \cite{baityjesi:18}, we obtain $t_3=e^{\beta E_1}$. Until $t_3$, $\emaxt$ typically still maintains its initial value.

\subsubsection{Aging}
\label{subsec:aging}
After $t_3$, $\emaxt$ starts increasing. We start by considering the aging stage, during which Eq.~\ref{eq:et} is valid.
We proceed analogously to Sec.~\ref{subsec:reachingcommonenergies} with an additional ingredient: high-energy records are harder to reach not only because transitions to them are rarely accepted -- as it was also previously --  but also because record-breaking states are rarely found among the neighbours of a given configuration. 

The probability that a trap has a neighbour of energy at least $E$ with probability $p_{\mathrm{find}}(E)\sim e^{-\frac{E^2}{2N}}$. Thus, when exiting a trap, a neighbour of energy $E$ is chosen over the lowest one with probability
\begin{equation}
\label{eq:ptransit}
p_{\mathrm{find}}(E)p_{\mathrm{accept}}(E)\sim e^{\beta(\inf I - E)-\frac{E^2}{2N}}\,.
\end{equation}
Then, in order to hit a new record, the number of visited traps should be 
\begin{equation}
\label{eq:ntptransit}
n_t\sim p_{\mathrm{find}}^{-1}(E)p_{\mathrm{accept}}^{-1}(E)\,.
\end{equation}

Again, by using extremum statistics (App.~\ref{subsec:asymptoticbounds}, Eq.~\eqref{eq:mn}), one has
\begin{equation}
\label{eq:nt}
n_t\sim e^{\frac{\emint^2}{2N}}
\end{equation}
Combining this with Eqs.~\eqref{eq:ptransit} and \eqref{eq:ntptransit} gives
\begin{equation}
\label{eq:firstfew1}
-\emint\approx\sqrt{2\beta N(\emaxt-\inf I)+\emaxt^2}\,.
\end{equation}
Plugging Eqs.~\eqref{eq:et} and \eqref{eq:Ieqiv} into Eq.~\eqref{eq:firstfew1}, we obtain
\begin{equation}
\label{eq:aging}
\emaxt=\beta N\left(-1+\sqrt{1-\frac{2}{\beta}\sqrt{\frac{2\ln N}{N}} +\left(\frac{\ln t}{\beta^2N}\right)^2}\right)\,.
\end{equation}
We can express Eq.~\eqref{eq:aging} in simpler forms by considering explicitly the different values of $\ln t$.

\paragraph{First records} We first consider the times for which $\ln t$ is of order $\beta E_1$ (but greater than it, so as to have $t>t_3$). Then, by substituting Eq.~\eqref{eq:e1}, Eq.~\eqref{eq:aging} becomes
\begin{equation}
\label{eq:firstfew2}
\emaxt=\sqrt{2N\ln N}\left(\left(\frac{\ln t}{\beta E_1}\right)^2-1\right)\,.
\end{equation}

\paragraph{Intermediate regime} We next assume that $\ln t$ is much larger than $\beta E_1$, but much smaller than $\beta^2 N$. Then Eq.~\eqref{eq:aging} takes the particularly simple form
\begin{equation}
\label{eq:intermediate}
\emaxt=\frac{\beta N}{2}\left(\frac{\ln t}{\beta^2 N}\right)^2\,.
\end{equation}

\paragraph{Exponential scales} Finally, we have $\ln t\sim \beta^2 N$. Then Eq.~\eqref{eq:aging} reduces to
\begin{equation}
\label{eq:longscales}
\emaxt=\beta N\left(-1 + \sqrt{1+\left(\frac{\ln t}{\beta^2 N}\right)^2}\right)\,.
\end{equation}

In order to determine time $t_4$ at which aging ends, we consider the mean energy, which is equal to $\emint$ as long as aging lasts. Then aging ends with $E_\MIN(t_4)=-\frac{\ln t_4}{\beta}$ equal to the equilibrium mean energy, itself given either by the global minimum 
of the energy or by $\langle E\rangle=\frac{\int Ee^{-\beta E}\rhogauss(E)\md E}{\int e^{-\beta E}\rhogauss(E)\md E}=-\beta N\,.$
Hence, by App.~\ref{subsec:asymptoticbounds}, \eqref{eq:globalmax}
\begin{equation}
\label{eq:t5}
t_4=e^{\beta\bi N}\,,
\end{equation}
where we introduced $\bi\equiv\min(\beta,\beta_c)$.

We stress that this regime describes the aging occurring at exponential time scales, that are generally regarded as the relevant ones for activated dynamics.
In this regime, $\emaxt$ {grows} as $\ln t$ (and so does $-\emint$) until the system has eventually 
equilibrated. Yet, since $\emaxt$ grows slower than $\emint$, it will keep increasing even after $\emint$ saturates.

\subsubsection{Equilibrium regime}
\label{subsec:verylongscales}
After $t_4$, $\emaxt$ continues to evolve even though the system has already reached macroscopic equilibrium. In this regime Eqs.~\eqref{eq:ptransit} and \eqref{eq:ntptransit} still hold, whereas Eq.~\eqref{eq:et} fails, making $\emint$ irrelevant for the calculation of $\emaxt$. 

The calculation that led to Eq.~\eqref{eq:t5} also gives that the mean time spent per state $\frac{t}{n_t}$ stations at the end of aging (the system reaches macroscopic equilibrium) at the value $\frac{t_4}{n_{t_4}}$. In other words, $\frac{t}{n_{t}}=\frac{t_4}{n_{t_4}}$ for all $t>t_4$.

Using Eqs.~\eqref{eq:ptransit}, \eqref{eq:ntptransit} and \eqref{eq:t5} together with the fact that $\inf I$ is negligible compared to $\emaxt$, we get
\[t=n_t\frac{t_4}{n_{t_4}}=n_te^{\beta\bi N}e^{-N\frac{\bi^2}{2}}=e^{\beta\bi N+\beta\emaxt-N\frac{\bi^2}{2}+\frac{\emaxt^2}{2N}}\,.\] Thus,
\begin{equation}
\label{eq:verylongscales}
\emaxt=\beta N\left(-1 +\sqrt{\left(1-\frac{\bi}{\beta}\right)^2+2\frac{\ln t}{\beta^2N}}\right)\,\,.
\end{equation}
Eq.~\eqref{eq:verylongscales} is valid until the global maximum $E_\MAX(\infty)=\beta_c N$ (see App.~\ref{subsec:asymptoticbounds}, Eq.~\eqref{eq:globalmax}) is reached. This means that it remains valid for
\begin{equation}\label{eq:verylongscales-ineq}
\ln t\leq (2\beta-\bi+\beta_c)(\beta_c+\bi)\frac{N}{2}= \begin{cases} N\frac{(\beta+\beta_c)^2}{2} & \textrm{ if } \beta<\beta_c\\
2N\beta_c\beta & \textrm{ if } \beta>\beta_c\,.
\end{cases}
\end{equation}
Inequality \eqref{eq:verylongscales-ineq} defines the typical time $t_5$ required to reach the global maximum
\begin{equation}
\label{eq:microeq}
t_5=e^{(2\beta-\bi+\beta_c)(\beta_c+\bi)\frac{N}{2}}\,.
\end{equation}

\subsubsection{Saturation}
\label{subsec:saturation}
Naturally, after $t_5$, that is after the global maximum has been attained, the maximum energy cannot change any more, so thereafter
\begin{equation}\label{eq:emax-tinfty}
\emaxt = \beta_cN\,.
\end{equation}

\subsection{Remarks}
\label{sec:comprehensive}
In our discussion, we treated the typical maximum of the energy $\emaxt$. We expect that when taking the thermodynamic limit there will be a concentration of the measure around a central value, so the same results should hold when replacing typical $\emaxt$ by the expectation $\mathbb{E}[{\emaxt}]$.
In App.~\ref{app:finitecorr}, we describe the subleading corrections that need to be taken into account in order to study finite systems, and tackle a few of them. The different regimes of $E(t)$ and $\emaxt$ (at leading order as $N\to \infty$) are summarised in Tab.~\ref{tab:summ}.

\begin{figure}[tb]\centering
\includegraphics[width=.45\textwidth]{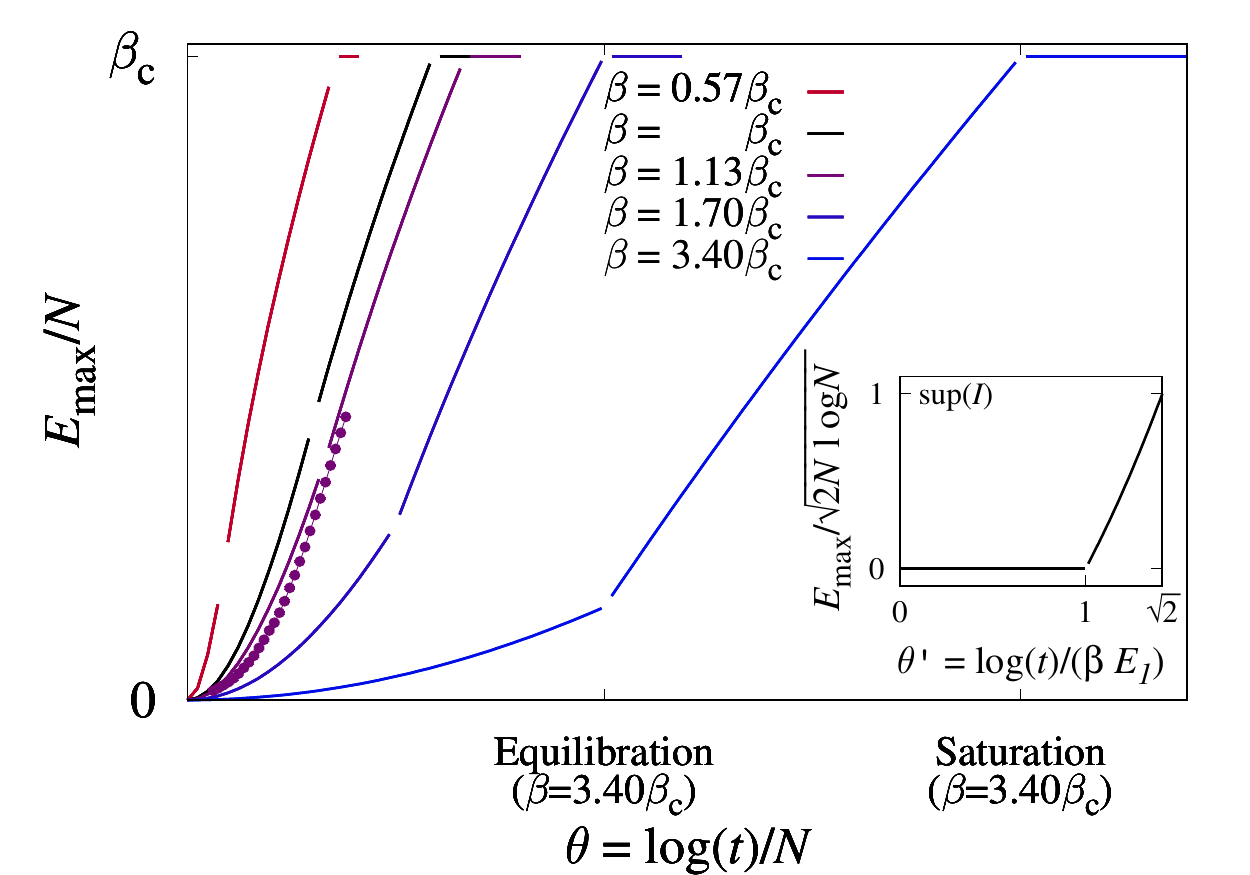}
\includegraphics[width=.45\textwidth]{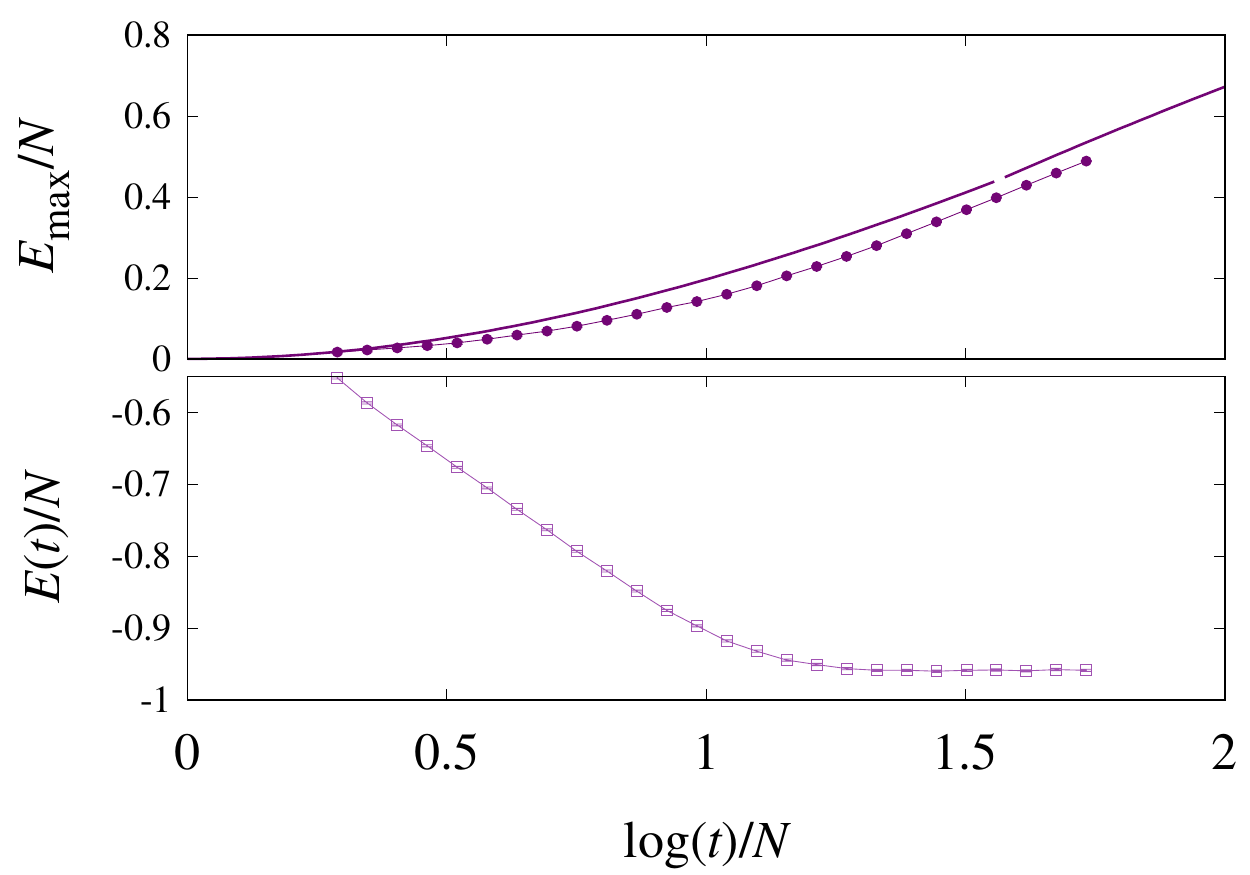}
\caption{Intensive maximum energy, $\emax$, in the \ac{rem} as a function of the rescaled time $\theta=\ln t/N$. \textbf{Left}: Plots for several values of the inverse temperature $\beta$. The $y$ axis is linear, in units of $\beta_c$. The curves are split in three, in order to stress the presence
of three subsequent regimes: Aging (Sec.~\ref{subsec:aging}), Equilibrium (Sec.~\ref{subsec:verylongscales}) and Saturation (Sec.~\ref{subsec:saturation}).
The tics on the $x$ axis show the beginning of these regimes for the highest $\beta$, corresponding to $T=0.25$.
We also show data from simulations at $\beta=1.13\beta_\mathrm{c}$, for systems of size $N=12$, which agrees well with our $N=\infty$ predictions. Finite-size corrections are discussed in App.~\ref{app:finitecorr}.
In the \textbf{inset} we show $\emax$ for shorter time scales $\theta'$ (see main text).
\textbf{Right}: The \textbf{top} figure depicts the same curves shown in the left panel, for $\beta=1.13\bc$. The solid line is the result of our calculation (in the thermodynamic limit), and the points are from runs with $N=12$ (error bars are smaller than the points).
The separation between the continuous lines emphasises the end of the aging regime according to our calculation [Eq.~\eqref{eq:t5}].
The \textbf{bottom} figure shows the energy as a function of time for comparison.\protect\\Numerical data here and in Fig.~\ref{fig:Emax_tw} are averaged over disorder realizations.
}
\label{fig:Maxbeta}
\end{figure}

\begin{table}[tb]
\begin{center}
\begin{tabular}{c|c|c|c}
Regime     & Time scale                          & $E(t)$/N           & $\emaxt/N$\\\hline\hline
Initial    & $t=0$                               & $\sim1/\sqrt{N}$   & $E(0)\sim1/\sqrt{N}$\\\hline
Drift      & $\ln t<\log_2 N$                    & \small{decreasing} & $E(0)\sim1/\sqrt{N}$\\\hline
First trap & $\ln t<\beta\sqrt{\frac{N}{\ln N}}$ & $\sqrt{2\ln N/N}$  & $E(0)\sim1/\sqrt{N}$\\\hline
Common	   & $\ln t<\beta E_1$                   & $-T\ln t/N$	      & $E(0)\sim1/\sqrt{N}$\\\hline
First records & $\ln t\sim\beta E_1$			 & $-T\ln t/N$	      & $\sqrt{\frac{2\ln N}{N}}\left(\left(\frac{\ln t}{\beta E_1}\right)^2-1\right)$\\\hline
Intermediate & $\beta E_1\ll\ln t\ll\beta^2 N$   & $-T\ln t/N$ & $\frac{\beta}{2}\left(\frac{\ln t}{\beta^2N}\right)^2$\\\hline
Exponential      & $\ln t<\beta\bi N$            & $-T\ln t/N$              &  $\beta\left(-1+\sqrt{1+\left(\frac{\ln t}{\beta^2 N}\right)^2}\right)$\\\hline
Equilibrium & $\ln t<(2\beta - \bi + \bc)(\bc+\bi)\frac{N}{2}$ & $-\bi$               & $\beta\left(-1+\sqrt{\left(1-\frac{\bi}{\beta}\right)^2+2\frac{\ln t}{\beta^2 N}}\right)$\\\hline
Saturation & any larger time scale & $-\bi$               & $\bc$\\\hline
\end{tabular}
\end{center}
\caption{Summary of the regimes of the dynamics evolution in the \ac{rem} in chronological order. 
For each regime (first column) we depict the time of validity, the typical intensive energy one finds at time $t$, $E(t)/N$, 
and the typical intensive record of the maximum energy, $\emaxt/N$.
The energy $E_1=\sqrt{2\beta N\sqrt{2N\ln N}}$ is defined in Eq.~\eqref{eq:e1}, $\bc=1/\Tc=\sqrt{2\ln 2}$ is the inverse critical temperature, and $\bi=\min(\bc,\beta)$.
}
\label{tab:summ}
\end{table}
Relevant time scales for the glassy activated dynamics stem naturally from our analysis. The longest time scale is exponentially large in the system size.
Thus, we can define a rescaled time $\theta=\ln t/N$, which is useful to express Eqs.~\eqref{eq:longscales}, \eqref{eq:verylongscales} and \eqref{eq:emax-tinfty} in a meaningful way. The intensive maximum energy for these time scales is plotted in Fig.~\ref{fig:Maxbeta}--left, with a comparison with numerics in small systems. 
In the thermodynamic limit, the piecewise concatenation of these regimes is continuous.\footnote{For $\beta<\bc$ its derivative is also continuous at $\theta=\beta\bi$, corresponding to $t_4$.} 

In units of $\theta$, all the previous time regimes collapse to zero in the thermodynamic limit.
For {the regime at which the first records are observed,} the relevant rescaled time is $\theta'=\ln t/(\beta E_1(N))$, whereas the energy should be rescaled by a factor $1/\sqrt{2N\ln N}$ [see Eq.~\eqref{eq:firstfew2}]. The resulting rescaled curve is plotted in the inset of Fig.~\ref{fig:Maxbeta}--left.

In Fig.~\ref{fig:Maxbeta}--right, we show that $\emaxt$ keeps growing even after the system has thermalised, and that the equilibration time, $t_4$, extracted from $\emaxt$ (in the thermodynamic limit) corresponds roughly to the time at which $E(t)$ reaches its equilibrium value (in a system of size $N=12$).

\section{Energy maxima and threshold energy in the \texorpdfstring{$\boldsymbol{p}$}{p}-spin model}\label{sec:pspin}

A natural question is how the behaviour of $\emax(t)$ extends to more complex glassy {systems}, and how it can be used to extract useful information on them.
We {consider} the $p$-spin model, since in the limit $p\to\infty$ the thermodynamics (though not the long-time activated dynamics~\cite{baityjesi:18c}) is the same as the one of the \ac{rem}.

In the $p$-spin model, there is a threshold energy $\Eth$ over which there are no energy minima~\cite{castellani:05}, 
as it also happens in the \ac{tm} and the \ac{rem}. Yet, at variance with the \ac{tm} and the \ac{rem}, where $\lim_{N\to\infty}(\Eth/N)=0$, in the $p$-spin model
$\Eth$ is extensively negative (i.e. $\lim_{N\to\infty}(\Eth/N)<0$).

Let us now take into account the maximum energy $\emax(t/2,t)$ reached in the time interval $[t/2,t]$.
Taking the maximum in a dilating time interval is useful to identify the separation of time scales that arises in the dynamics of glassy systems. In the thermodynamic limit, the dynamics of the $p$-spin model on time scales not diverging with $N$ is known \cite{cugliandolo:93}: the energy decreases monotonically towards the threshold energy. Moreover, in this regime, the energy as a function of time has fluctuations of order $\sqrt{N}$. In consequence, $\emax(t/2,t)$ is equal to the energy at time $t/2$ and approaches $\Eth$ for $t$ large but not diverging with $N$. On times diverging with $N$ the dynamics becomes activated (on time-scales exponentially large in $N$), the intensive energy goes below $\Eth$, and the energy starts to have rare high excursions that increase the value of $\emax(t/2,t)$ (see also Fig.~\ref{fig:sample} in App.~\ref{sec:sample}). Hence, one can expect that $\emax(t/2,t)$ grows close to logarithmically, as in record-breaking dynamics for i.i.d. random variables (and as it happens for $\emaxt$ in the \ac{rem} during the aging regime, see table \ref{tab:summ}).

\begin{figure}[tb]
\centering
\includegraphics[width=0.49\columnwidth]{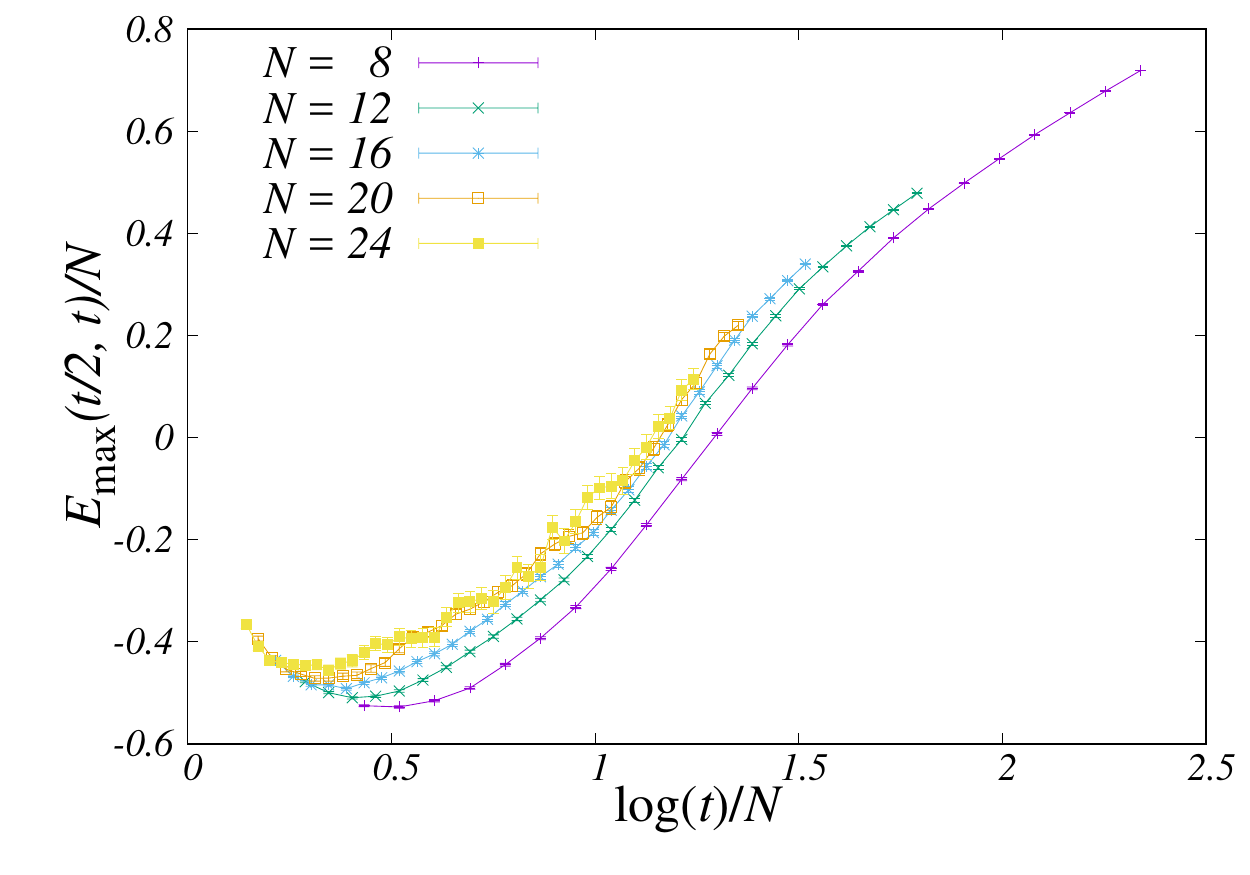}
 \resizebox{0.49\textwidth}{!}{\input{Emax_tw_tex}}
\caption{
Intensive maximum energy $E_\MAX(t/2,t)/N$ reached in the time interval $[t/2,t]$, for different system sizes $N$. 
On the \textbf{left}, we show data for the \ac{rem} at $T=0.75$. Decreasing 
the temperature (not shown), the bottom of the curve becomes progressively flatter,
but the lowest point stays approximately at the same height.
On the \textbf{right}, we show data for the $3$-spin model at $T=0.54$.
In both cases the curves seem to converge to a master curve proportional to $\ln t$. \textbf{Inset}: Scaling behaviour of the size-dependent threshold energy $E_{\text{th}}(N)$ (defined as the minimum of $\emaxt$) in the $3$-spin model. 
The prediction from Ref.~\cite{rizzo:13}, valid for infinitely large systems, $\Eth=-0.762$, is also shown (red square). 
The dotted line represents the fit $E_{\text{th}}(N)=aN^{-1/2}+b$. 
The coefficients of the fit are $a=1.47(1)$ and $b=-0.758(1)$.
}
\label{fig:Emax_tw}
\end{figure}

The minimum value reached by $\emax(t/2,t)$ can be used to {define} the value of $\Eth$ in numerical simulations. This method adds to the usual procedures of calculating the threshold, and has the advantage of not requiring extrapolations to infinite times~\cite{cugliandolo:93}, nor the computation of the complexity of minima~\cite{crisanti:95}.
In Fig.~\ref{fig:Emax_tw} we show $\emax(t/2,t)$ both in the \ac{rem} and in the $p$-spin model (with $p=3$), for different system sizes.
As expected, in both cases the curve reaches a minimum which we identify as $\Eth$. Note that, in the \ac{rem}, 
$\frac{\Eth}{N}$ grows as $N$ increases whereas in the $3$-spin model it decreases, in agreement with the fact that 
the intensive threshold energy is zero in the former model and negative in the latter. 
In the inset of Fig.~\ref{fig:Emax_tw}, we show that, in the $p$-spin model, the finite-size threshold energy obtained through this procedure is controlled by the $N^{-1/2}$ fluctuations, and converges to its analytical value in the thermodynamic limit~\cite{rizzo:13}.\footnote{This same scaling is not as clean in the REM, probably because the system sizes are too small. In App.~\ref{app:finitecorr} we argue that the asymptotic limit is reached for $N\sim10^3$.}
From the figure one can also see that, as argued in the previous paragraph, the relevant time scales for the growth of $\emax(t/2,t)$ are exponential.

\section{Conclusions}\label{sec:conclusions}

By counting the number of visited traps, and noting that in \ac{tm}-like models almost all the time is spent in the deepest trap, we were able to calculate the evolution of the maximum energy reached after a time $t$, $\emaxt$. 
Our method, which applies record statistics to glasses, allowed us to add to the very reduced number of analytical calculations of activation in glasses (see also~\cite{bryngelson:89,gayrard:18}). It has the advantage of transparently treating the dynamics, providing new insight on the long-time behaviour of glassy systems. Further, we showed how the behaviour of the maximum energy record naturally reveals the different time scales involved in glassy relaxation, and how it can be used to identify a crucial static quantity, the threshold energy $\Eth$.
Calculating $\Eth$ by identifying the minimum of $\emax(t/2,t)$ is arguably simpler than calculating the complexity of the minima~\cite{crisanti:95}, and does not require any kind of extrapolations to infinite times as required by traditional methods~\cite{cugliandolo:93}.

The relationship between $\Eth$ and $\emax(t/2,t)$ relies on the assumption that it is necessary to reach $\Eth$ to connect low-lying energy minima. Therefore, its success in the $p$-spin model is a strong suggestion that there are no low-energy paths connecting minima in the $p$-spin.

Our results also connect to recent phenomenological works, suggesting to rationalise the dynamical slowdown of low-temperature
glasses as a record-breaking process over larger and larger domains~\cite{boettcher:05b, boettcher:11, robe:16}.
Our mean-field calculations could then account for this process in regions smaller than the correlation length, where the system is effectively fully-connected.

\section*{Acknowledgements}
We thank Valerio Astuti, Chiara Cammarota, Claude Godr\`eche and Satya Majumdar for interesting discussions.
This work was funded by the Simons Foundation for the collaboration ``Cracking the Glass Problem" (No. 454935 to G. Biroli and No. 454951 to D.R. Reichman). 
M.B.-J. was partially supported through Grant No. FIS2015-65078-C2-1-P, jointly funded by MINECO (Spain) and
FEDER (European Union).

\clearpage
\appendix
\section{Typical neighbours' energies}
\label{app:I}
In this section we recall some useful results from classical extremum statistics that we use thorough in our paper. For a more detailed discussion
see, for example, Refs.~\cite{gumbel:58,schehr:13}.

\subsection{Asymptotic bounds}
\label{subsec:asymptoticbounds}
We would like to determine the typical value of the maximum, $M_n$, out of $n$ \ac{iid} random variables drawn from distribution $\rhogauss(E)$
[Eq.~\eqref{eq:rhogauss}]. We do this for arbitrary $n=n(N)\to\infty$, as different applications will be needed.

Let $X$ be a random variable drawn from the distribution $\rhogauss$. Its cumulative distribution $F(x)$ is the probability that $X$ is smaller than $x$, $F(x)=\mathbb{P}(X\leq x)$.
We consider \ac{iid} random variables, so the probability that the maximum of $n$ tries is smaller than some constant $a$ is
$\mathbb{P}(M_n\leq a)=(F(a))^n$.

We can find a location-scale transformation to show that $M_n$ asymptotically follows the Gumbel distribution $\mathscr{L}_{\mathrm{GUM}}((-\infty,x])=e^{-e^{-x}}$.
\footnote{Mind that the Gumbel distribution is not centred: its mean is the Euler-Mascheroni constant $\gamma$.}
Set 
\begin{equation}\label{eq:anbn}
a_n=\sqrt{\frac{N}{2\ln n}},\,b_n=\sqrt{2N\ln n}\left(1-\frac{\ln\ln n+ \ln (4\pi)}{4\ln n}\right)\,.
\end{equation}
Then integrating by parts we get
\begin{multline}
\label{eq:1-F}
1-F(b_n+xa_n)=\frac{1}{\sqrt{2\pi N}}\int_{b_n+xa_n}^\infty e^{-\frac{y^2}{2N}}\md y\stackrel{n\rightarrow\infty}{\sim}\sqrt{\frac{N}{2\pi}}e^{-\frac{(b_n+xa_n)^2}{2N}}\frac{1}{b_n+xa_n}\\\sim\frac{1}{\sqrt{4\pi\ln n}}\exp\left(-x-\frac{b_n^2}{2N}\right)\sim\frac{e^{-x}}{n}\,.\hfill
\end{multline}
By isolating $F(b_n+xa_n)$ and recalling that $\mathbb{P}(M_n\leq a)=(F(a))^n$, one gets that
\begin{equation}
\label{eq:GUMconv}
\mathbb{P}\left(\frac{M_n-b_n}{a_n}\leq x\right)\stackrel{n\to\infty}{\longrightarrow}\mathscr{L}_\mathrm{GUM}\,.
\end{equation}

\paragraph{Explicit cases}
The two particular cases relevant for our purposes are $n=2^N$ and $n=N$, which respectively indicate the global energy maximum of the system (there are $2^N$ sites in the system), and the highest-energy neighbour (each site has $N$ neighbours). 
For the global maximum we inject $n=2^N$ in the expression~\eqref{eq:anbn} of $b_n$ and obtain
\begin{equation}
\label{eq:globalmax}
E_\MAX(\infty) = -E_\MIN(\infty) = M_{2^N} = \bc N+\mathcal{O}(\ln N)\,.
\end{equation}

In the second case, $n=N$, we have
\begin{equation}
\label{eq:Iasymp}
-\inf I = \sup I = M_N = \sqrt{2N\ln N}\left(1+\mathcal{O}\left(\frac{\ln\ln N}{\ln N}\right)\right)\,.
\end{equation}

Moreover, to the leading order Eq.~\eqref{eq:GUMconv} gives
\begin{equation}
\label{eq:ne}
n\approx e^{\frac{M_n^2}{2N}}
\end{equation}
or equivalently
\begin{equation}
\label{eq:mn}
M_n\approx\sqrt{2N\ln n}\,.
\end{equation}
It is worth noting that, according to Eq.~\eqref{eq:GUMconv}, the energy difference between a local minimum and its lowest neighbour is 
of order $a_N$ (see also \cite[Sec.~3.1.2]{schehr:13}). 

Finally, note that due to the concentration of $M_n$ around $b_n$, for large systems one can treat both the global maximum of the energies and $I$ as deterministic.

\label{subsec:lowestneighboursgap}

\subsection[Finite \texorpdfstring{$N$}{N}]{Finite \texorpdfstring{$\boldsymbol{N}$}{N}}
\label{subsec:finiten}
When dealing with small systems (see App.~\ref{app:finitecorr}) we need to go beyond the asymptotic bounds. 
To get the exact expression for $\sup(I)=M_N$ we can use again the fact that 
$\mathbb{P}(M_n\leq x)=(F(x))^n$. From there we can extract the distribution of $M_n$, and 
calculate its average, which results in
\begin{equation}
\label{eq:boundsI}\sup{I}=-\inf{I}= \sqrt{\frac{N}{2\pi}}\int_{-\infty}^{+\infty} Nx(1-\erfc(x))^{N-1}e^{-\frac{x^2}{2}}\md x\,,
\end{equation}
where
\begin{equation}
\label{eq:erfc}
\erfc(x)=\frac{1}{\sqrt{2\pi}}\int_x^\infty e^{-\frac{u^2}{2}}\md u\,.
\end{equation}
For example, for $N=20$, this gives $\sup I\approx 8.34$.

\section{Corrections for small systems}
\label{app:finitecorr}
In this appendix we take into account subleading terms of $\emaxt$, so as to make quantitative predictions also for small system sizes.
The general reasoning will remain the same as in Sec.~\ref{sec:rem}, but we will pay particular attention to the fact that $-\inf I$ is not much smaller than $N$ and we will no longer use the asymptotic value of $\inf I$, in favour of Eq.~\eqref{eq:boundsI}.

\subsection{The initial condition}
It is convenient to study not only $\mathbb{E}[\emaxt]$ -- the expectation over different instances of the \ac{rem} -- but also the expectation given the starting energy 
$\mathbb{E}[\emaxt|E(0)]$.
A particular case that will be useful in the following, is when the dynamics starts at the threshold energy,
\begin{equation}
\mathcal{E}(t):=\mathbb{E}[\emaxt|E(0)\approx\inf I] \,.
\end{equation}
From $\mathcal{E}(t)$ we can compute $\emaxt$ for any starting condition. If $E(0)<\inf I$, we have 
\begin{equation}
\label{eq:delayed2original}
\mathbb{E}[\emaxt|E(0)]=\begin{cases}
E(0) & \mathrm{ if }\; t\ll e^{\beta(\inf I-E(0))}\\
\mathcal{E}(t-e^{\beta(\inf I-E(0))})\approx \mathcal{E}(t) & \mathrm{ if }\; t\gg e^{\beta(\inf I-E(0))}\,,
\end{cases}
\end{equation}
whereas if $E(0)>\inf I$ and $t\gg N$, we get
\begin{equation}
 \mathbb{E}[E_{\MAX}(t)|E(0)]=\max(E(0),\mathcal{E}(t))\,.
\end{equation}

$\mathcal{E}(0)$ can be regarded as the energy typically reached after at time of order $t_1=N$ (see Sec.~\ref{sec:rem}). Hence, studying $\mathcal{E}(t)$ is equivalent to looking at $\mathbb{E}[\emax(t_1,t)]$, so disregarding 
the initial part of the dynamics (until reaching the first local minimum). This time delay is small compared to the other time scales in play.

\subsection{Corrections}
\subsubsection{General corrections}
The main finite-size corrections we treat are the following:
\begin{itemize}
 \item Finite-size corrections to extreme value results.
 \item Typical depth of traps.
 \item Alternative paths for reaching high energies.
\end{itemize}
Further corrections that we do not take into account are stressed in Sec.~\ref{subsec:remaining-errors}.

\paragraph{Finer analysis of traps and their barriers}
Consider a local minimum of the energy landscape and denote its energy by $E$. 
Then $E$ is simply the minimum of $N+1$ energies from $\rhogauss$, which is close to $\inf I$. 
However, attention should be paid to the energy of its lowest neighbour, which is not simply $\inf I$,
since it is conditioned to be at energy higher than $E$. The average energy $\phi(E)$ of the lowest neighbour of a configuration with energy $E$ should then be rewritten as
\begin{equation}
\label{eq:defphi}
\phi(E)=\frac{\sqrt{N}\int_{\frac{E}{\sqrt{N}}}^\infty x(1-\erfc(x))^{N-1}e^{-\frac{x^2}{2}}\md x}{\int_{\frac{E}{\sqrt{N}}}^\infty (1-\erfc(x))^{N-1}e^{-\frac{x^2}{2}}\md x}\,.
\end{equation}
The function $\phi(E)$ goes down from an initial value a bit higher than $\inf I$, when $E=\inf I$, to $\inf I$, when $E$ is a few standard deviations of $\inf I$ lower than it. 
In particular, by the end of aging we can safely write $\phi(\mathcal{E}_{\min}(t))=\inf I$, since the system will
have reached the lowest-energy configuration.

Using Eq.~\eqref{eq:defphi}, the exit time of a trap,
that corresponds approximately to the mean time required to go to the lowest neighbour,
is rewritten as
\begin{equation}\label{eq:exittime}
t_\mathrm{exit}=Ne^{\beta\left(\phi(E)-E\right)}\,.
\end{equation}

\paragraph{Aging regime}
During the aging stage, since most of the time is spent in the lowest-energy state visited so far, the total elapsed time can be expressed as
\begin{equation}\label{eq:corraging}
t\approx t_\mathrm{exit}^\mathrm{(aging)}\approx Ne^{\beta\left[\phi\left(\mathcal{E}_{\min}(t)\right) - \mathcal{E}_{\min}(t)\right]}\,.
\end{equation}
Hence, taking into account the relations in App.~\ref{subsec:asymptoticbounds}, the aging now ends after a time of order
\begin{equation}
\label{eq:endaging6} 
t_4'=Ne^{\beta(\inf I + \bi N)}\left(\sqrt{2\pi N}\bi e^{-\gamma}\right)^{-\frac{\beta}{\bi}}\,,
\end{equation}
where we also centred the Gumbel distribution, which gives the $\gamma$ term, and used $\phi(\mathcal{E}_{\min}(t'_4))\approx \inf I$. Retaining just the main order gives
\begin{equation}
\label{eq:endaging6main}
t_4^{(0)}=e^{\beta(\inf I+\bi N)}\,.
\end{equation}

\subparagraph{Time at lowest neighbours}
The time spent at lowest neighbours of traps is about $Nn_t$, since, due to the Metropolis rate 
\eqref{eq:mc}, the time spent before leaving them is of order $N$. Recall that if one draws $Nn_t$ energies 
from $\rhogauss$, their minimum is distributed as the minimum of $n_t$ lowest neighbours 
(see also Sec.~\ref{subsec:reachingcommonenergies}). 
Then, according to Eqs.~\eqref{eq:anbn} and~\eqref{eq:GUMconv} in Sec.~\ref{subsec:asymptoticbounds}, 
defining $y=\ln(Nn_t)$, one has
\[\mathcal{E}_\MIN(t)=\sqrt{2Ny}\left(1-\frac{\ln y+\ln(4\pi)-2\gamma}{4y}\right)\,.\]
The second term on the right hand side is small, so it can be treated as a perturbation. 
To the first order, one obtains
\begin{equation}
\label{eq:Nnt}
\ln(Nn_t)=\frac{\mathcal{E}_\MIN^2(t)}{2N}\left(1+N\frac{\ln\frac{\mathcal{E}_\MIN^2(t)}{2N} + \ln(4\pi)-2\gamma}{\mathcal{E}_\MIN^2(t)}\right)\,.
\end{equation}

\subparagraph{Records via other neighbours}
As most of the time is spent at minima, a new record $E$ for the maximum energy can be attained either transiting directly from the minimum, $E_\mathrm{trap}$, or via one of its neighbours. Let us denote by $A(E)$ the probability that $E$ is reached from the trap or its neighbours divided by the probability that it is reached directly from the trap.

The probability of reaching $E$ directly from the trap is given by the time spent there,
\[N\exp(\beta(\phi(E_{\mathrm{trap}})-E_\mathrm{trap})),\] 
times the transition rate $\exp(\beta(E_\mathrm{trap}-E))/N$. Let us consider the lowest neighbour, $\phi(E_\mathrm{trap})$, which typically has two neighbours (including the trap) with lower energy. It has exactly half the transition probability to $E$ as directly from the trap, because the less time spent in it, $N/2$, cancels out with the higher transition rate towards $E$. The second-lowest energy has typically three lower neighbours, so by the same reasoning it contributes as a third of the direct transition from the trap. Continuing this way we get a total 
transition rate of $H_n$ times the one directly from the trap, where $H_n$ is the harmonic partial sum.

However, this reasoning only concerns the neighbours {of the trap with} energies lower than $E$. In the general case we get 
 \begin{equation}
 \label{eq:A}
 A(E)\approx \max\left(\ln\left(N\erfc\left(\frac{-E}{\sqrt{N}}\right)\right),1\right)\,,
 \end{equation}
where we approximate $H_n$ by $\ln n$ and recall the fact that transiting directly from the trap is always possible to avoid errors for $E$ close to $\inf I$.

\subsubsection{Delayed maximum}
\paragraph{Long-time aging regime}
Note that most of the lowest neighbours of visited traps are at energy $\phi(\inf I)$. 
Then, due to the previous observations $\mathcal{E}(t)$ should satisfy, analogously to Sec.~\ref{subsec:aging},
\begin{equation}
\label{eq:aginggros}
\begin{split}
1&\simeq Nn_t \int_{\mathcal{E}(t)}^\infty A(E)q_{\phi(\inf I),E}N\rhogauss(E)\md E\\
&\approx e^{-\gamma}\sqrt{\frac{2\pi\mathcal{E}_\MIN^2(t)}{N}}e^{\frac{\mathcal{E}_{\min}^2(t)}{2N}}A(\mathcal{E}(t))\int_{{\mathcal{E}(t)}}^\infty \frac{1}{N}e^{\beta(\phi(\inf I)- E)}N\rhogauss(E)\md E\\
&\approx e^{-\gamma}A(\mathcal{E}(t))\frac{\sqrt{N}|\mathcal{E}_\MIN(t)|}{(\mathcal{E}(t)+\beta N)\sqrt{2\pi}}\sqrt{\frac{2\pi}{N}} e^{\frac{\mathcal{E}_{\min}^2(t)}{2N}}e^{\beta (\phi(\inf I)-\mathcal{E}(t))}e^{-\frac{\mathcal{E}^2(t)}{2N}}\\
&\approx e^{-\gamma}B(\mathcal{E}(t))\frac{|\mathcal{E}_\MIN(t)|}{\beta N}e^{\frac{\mathcal{E}_{\min}^2(t)}{2N}}e^{\beta (\phi(\inf I)-\mathcal{E}(t))}e^{-\frac{\mathcal{E}^2(t)}{2N}}\,,
\end{split}
\end{equation}
where $q_{i,j}$ is the Metropolis rate from state $i$ to state $j$, and in the last step we set
\begin{equation}
\label{eq:B}
B(E):=\frac{A(E)}{1+\frac{E}{\beta N}}\,.
\end{equation}
Notice that the subexponential terms are only very small perturbations, so one can still solve the transcendental equation. Then, combining \eqref{eq:corraging} and \eqref{eq:aginggros} yields to the leading order
\begin{equation}
\label{eq:longscales6main}
\mathcal{E}^{(0)}(t)\approx-\beta N+ \sqrt{(\beta N+\inf I)^2 - 2\frac{1}{\beta}\inf I\ln t + \left(\frac{1}{\beta}\ln t\right)^2}\,.
\end{equation}
At this point, one can use Eqs.~\eqref{eq:B} and~\eqref{eq:longscales6main} in the bottom line of Eq.~\eqref{eq:aginggros}, to approximate the solution recursively, obtaining
\begin{equation}
\label{eq:longscales6}
\begin{split}
\mathcal{E}(t)\approx&-\beta N+ \Bigg(\beta^2 N^2 +2\beta N \phi(\inf I)+\left(\mathcal{E}_\MIN(t)\right)^2\\
&{+2N\ln\frac{\mathcal{E}_\MIN(t)}{\beta N}}+2N\ln B\left(\mathcal{E}^{(0)}(t)\right)-2N\gamma\Bigg)^{\frac{1}{2}}\,.
\end{split}
\end{equation}
where we used that $\phi(\mathcal{E}_\mathrm{min})=\inf I$, so this expression can only be valid deep in the aging regime.

\paragraph{Equilibrium regime} 
After the end of aging ($t>t_4'$, Eq.~\eqref{eq:endaging6}) we use that the mean time spent per state stations at its value at the end of aging $\frac{t_4'}{n_{t_4'}}$. 
{Also, at the end of aging one has $Nn_{t_4'}\approx e^{\frac{N\bi^2}{2}}$. Hence, following Eq.~\eqref{eq:endaging6},}
\begin{equation}
\label{eq:postaging}
\begin{split}
1\simeq&\, t\frac{Nn_{t_4'}}{t_4'}\int_{\mathcal{E}(t)}^\infty A(E)q_{{\phi(\inf I)},E} N\rhogauss(E)\md E\approx\\
\approx&\, \frac{t}{N}e^{-\beta(\bi N+\inf I)}\left(\sqrt{2\pi N}\bi e^{-\gamma}\right)^{\frac{\beta}{\bi}} e^{\frac{N\beta_\mathrm{i}^2}{2}}A(\mathcal{E}(t))\cdot\int_{\mathcal{E}(t)}^\infty e^{\beta({\phi(\inf I)} - E)}\rhogauss(E)\md E\approx\\
\approx&\, \frac{\sqrt{N}A(\mathcal{E}(t))}{(\mathcal{E}(t)+\beta N)\sqrt{2\pi}}\left(\sqrt{2\pi N}\bi e^{-\gamma}\right)^{\frac{\beta}{\bi}} \frac{t}{N}e^{-\beta\bi N}e^{\frac{N\beta_\mathrm{i}^2}{2}}\cdot e^{-\beta \mathcal{E}(t)}e^{-\frac{\mathcal{E}^2(t)}{2N}}{e^{\beta(\phi(\inf I)-\inf I)}}\approx\\
\approx&\, C(\mathcal{E}(t))\left(\sqrt{2\pi N}\bi\right)^{\frac{\beta}{\bi}-1}e^{-\frac{\gamma\beta}{\bi}}\frac{t}{N}e^{-\beta\bi N}e^{\frac{N\beta_\mathrm{i}^2}{2}} e^{-\beta \mathcal{E}(t)}e^{-\frac{\mathcal{E}^2(t)}{2N}}{e^{\beta(\phi(\inf I)-\inf I)}}\,,
\end{split}
\end{equation}
where in the last step we set 
\begin{equation}
\label{eq:C}
C(E):=A(E)\frac{\bi}{\beta+\frac{E}{N}}\,.
\end{equation}

Like in the previous paragraph, neglecting subexponential factors and solving \eqref{eq:postaging} for $\mathcal{E}(t)$ gives the main order
\begin{equation}
\label{eq:verylongscales6main}
\frac{\mathcal{E}^{(0)}(t)}{N}=-\beta+\sqrt{(\beta-\beta_\mathrm{i})^2 + \frac{2}{N}\ln t}\,,
\end{equation}
which is (accidentally) the same as \eqref{eq:verylongscales}. Then the final result is
\begin{equation}
\label{eq:verylongscales6}
\begin{split}
\frac{\mathcal{E}(t)}{N}=&-\beta+\Bigg((\beta-\beta_\mathrm{i})^2 + \frac{2}{N}\ln \frac{t}{N}+\frac{2}{N}\ln C\left(\mathcal{E}^{(0)}(t)\right)\\&+\frac{\beta-\bi}{N\bi}\ln\left(2\pi N \bi^2\right)
-\frac{2\gamma\beta}{N\bi}{+2\frac{\beta}{N}(\phi(\inf I) - \inf I)}\Bigg)^{\frac{1}{2}}\,.
\end{split}
\end{equation}

We should note that the piecewise concatenation of these functions is not expected to give good results around $t_4'$. That is because the right hand side of \eqref{eq:postaging} should actually be the same replacing $t$ by $t-t_4'$ and adding the right hand side of \eqref{eq:aginggros}. However, the transition between the two expressions happens very quickly -- just within a few times $t_4'$, which is nearly invisible on the semi-logarithmic plot. What is more, the expression should be further smoothed by the randomness of $t_4'$, which we took for constant.

\subsubsection{Original maximum}
Now that we know $\mathcal{E}(t)$, we can use it to deduce $\mathbb{E}[\emaxt|E(0)]$ and then obtain $\mathbb{E}[\emaxt]$ by {integrating over} $E(0)$. Using \eqref{eq:delayed2original} we get
\begin{equation*}
\begin{split}
\mathbb{E}[\emaxt]=&\mathbb{E}[\mathbb{E}[\emaxt|E(0)]]\approx\\
\approx&\int_{\inf I-T\ln t}^\infty \max(\mathcal{E}(t),E)\rhogauss(E)\md E+\int_{-\infty}^{\inf I-T\ln t} E\rhogauss(E)\md E=\\
=&\left(\erfc\left(\frac{\inf I-T\ln t}{\sqrt{N}}\right)-\erfc\left(\frac{\mathcal{E}(t)}{\sqrt{N}}\right)\right)\mathcal{E}(t)+\int_{|\mathcal{E}(t)|}^{T\ln t-\inf I} E\rhogauss (E)\md E
\approx\\
\approx&\left(1-\erfc\left(\frac{\mathcal{E}(t)}{\sqrt{N}}\right)\right)\mathcal{E}(t)+\sqrt{\frac{N}{2\pi}}e^{-\frac{\mathcal{E}(t)^2}{2N}}\,,
\end{split}
\end{equation*}
i.e.
\begin{equation}
\label{eq:max06}
\mathbb{E}[\emaxt]\approx\left(1-\erfc\left(\frac{\mathcal{E}(t)}{\sqrt{N}}\right)\right)\mathcal{E}(t)+\sqrt{\frac{N}{2\pi}}e^{-\frac{\mathcal{E}(t)^2}{2N}}\,,
\end{equation}which can be made explicit by plugging \eqref{eq:endaging6}, \eqref{eq:longscales6} and \eqref{eq:verylongscales6} in it. This expression approaches $\mathcal{E}(t)$ for large $t$ and remains small as long as $\mathcal{E}(t)<0$.

\subsection{Outlook on the finite-\texorpdfstring{$N$}{N} corrections}
\label{subsec:remaining-errors}
We analysed preasymptotic corrections that should be taken into account in order to compare our predictions
with numerical simulations. 
Because these corrections come from different sources, it was not possible quantify their magnitude as some
order of $N$. As we show at the end of this section, the corrections we accounted for are sufficient for a good
match with numerical simulations. Still, it should be noted that there are several effects which we did not account
for.

\emph{Difference between $\emint$ and $E(t)$.}
An important source of error are the deviations from the theory of $\emint$, which were observed in \cite{baityjesi:18}.
Since our calculations are based on $\emint$, those are likely to have an impact here, too.

\emph{Asymptotic statistics.}
Eqs.~\eqref{eq:aginggros} and \eqref{eq:postaging} are only valid up to a factor of order 1 on the left hand side. 
Moreover, we arbitrarily used mean values for both $\inf I$ and the Gumbel distribution, which should give errors on the effective value adapted 
to our observable of the order of their standard deviations. 

\emph{Time shift.}
The correction introduced by using $\phi$ [Eq.~\eqref{eq:defphi}] should also slightly impact the end 
of aging [Eq.~\eqref{eq:endaging6}] and thereby the whole post-aging regime. The same holds for the use of the asymptotic relation $\emin(\infty)\approx-\bc N$.

\emph{Returns.}
By considering the energies without memory of the trajectory's history, we disregarded returns to recently visited traps, which dominate the dynamics at subexponential time scales \cite{baityjesi:18}.
Those should effectively increase the exit times from traps. 

\begin{figure}[htb]
\includegraphics[width=.49\textwidth]{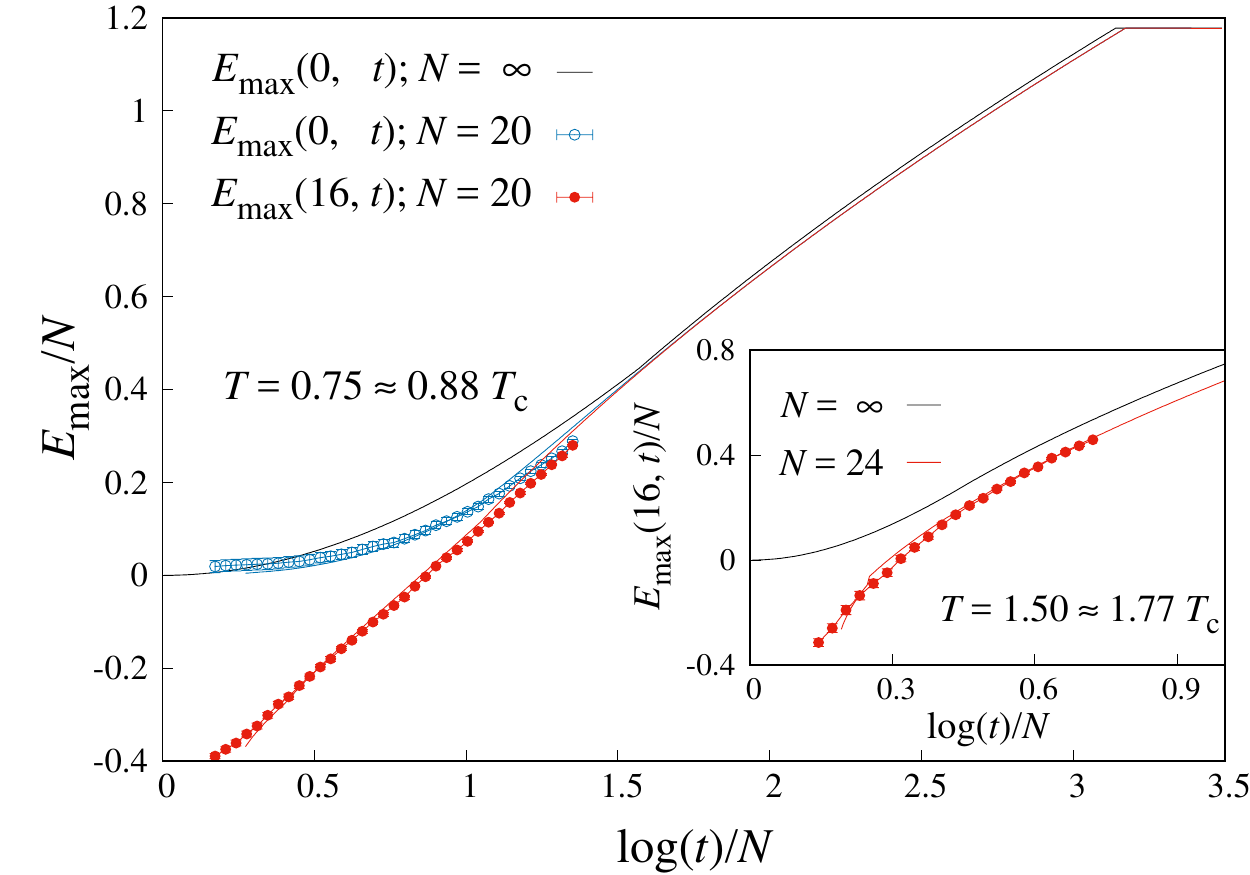}
\includegraphics[width=.49\textwidth]{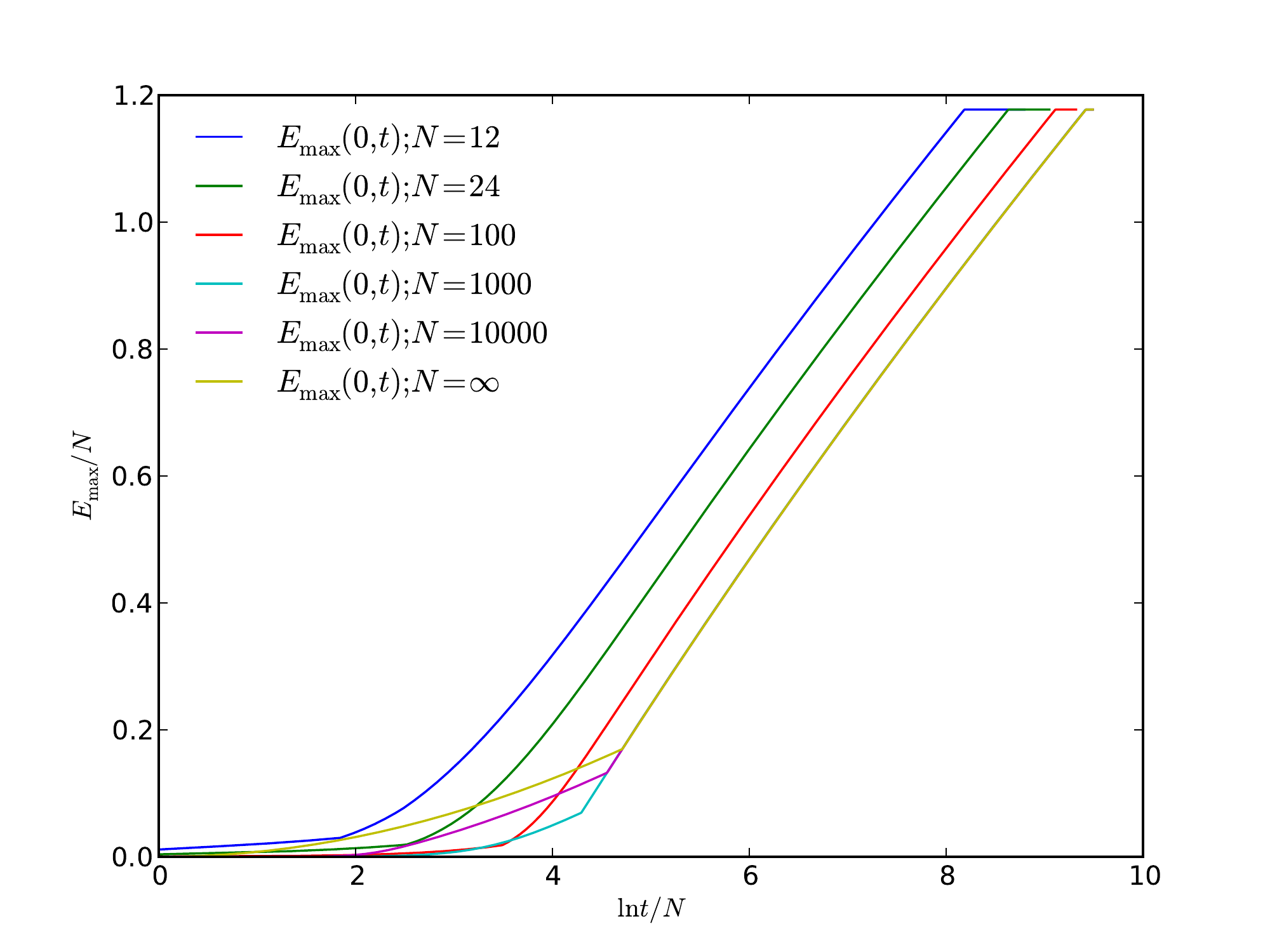}
\caption{Finite-size predictions for the intensive maximum energy, $\emaxt/N$, in the \ac{rem} as a function of the rescaled time $\theta=\ln t/N$. \textbf{Left: }Comparison between asymptotic solution, finite-size predictions and numerical simulations.
 We show analytical curve and numerical points for both $\emax(0,t)$ and $\emax(16,t)$. We show $T=0.75<\Tc$ in the main figure, 
 and $T=1.50>\Tc$ in the \textbf{inset}. \textbf{Right: }Convergence of the finite-size predictions for $\emax(0,t)$ 
 to the asymptotic solution at $T=0.25\approx 0.29\Tc$. Note that the convergence is non-monotone and very slow, as we are deep in the glassy state. Moreover, the convergence after the equilibration time is faster.}
\label{fig:finite-size}
\end{figure}

By comparing with numerical simulations (Fig.~\ref{fig:finite-size}--left), 
one sees that the effects we neglected seem sufficiently unimportant. 
Theoretical and numerical curves follow the same trend without using any free parameter.

\paragraph{Convergence to the asymptotic behaviour}
In figure Fig.~\ref{fig:finite-size}--right we show the convergence of the finite-size $\emaxt$ to its asymptotic limit,
which is different depending on the regime. For long times, $\emaxt/N$ becomes smaller with increasing $N$, and saturates
at around $N=1000$.
At short times, the trend is inverted, and finite-size effects are sizeable for even larger system sizes.
This slow speed of convergence explains why it is necessary to take into account finite-size corrections when 
comparing to numerical simulations.

\section{Records in a single trajectory}\label{sec:sample}
Discussing the evolution of the maximum in a single trajectory can simplify the concepts that we exposed throughout the article.
\begin{figure}[!htb]
\centering
\includegraphics[width=0.7\columnwidth]{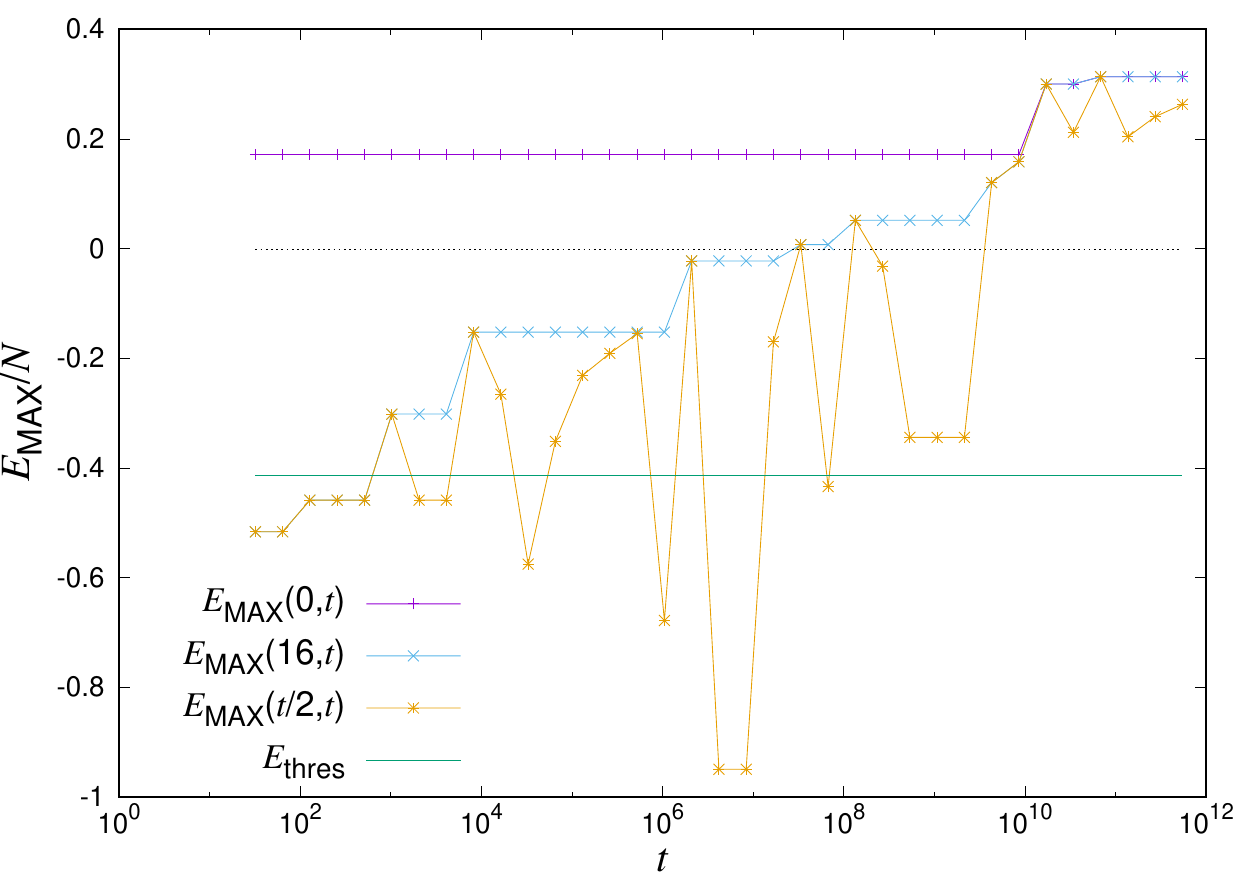}
\caption{A random sample of the \ac{rem} Metropolis dynamics, with $N=20, T=0.75$. We show the maximal attained energy, $E_\MAX(0,t)$, 
the one disregarding the first 16 time steps, $E_\MAX(16,t)$, and the one between $\frac{t}{2}$ and $t$, $E_\MAX(t/2,t)$. 
The green horizontal line represents the threshold energy $\Eth$ given by Eq.~\eqref{eq:boundsI}. The black dashed horizontal line is a reference for $E=0$.
}
\label{fig:sample}
\end{figure}
In Fig.~\ref{fig:sample} we show the evolution of the maximum energy reached in a single trajectory. We use three different indicators:
$E_\MAX(0,t)$ is the maximum energy since the beginning of the run, 
$E_\MAX(16,t)$ is a delayed maximum in which we wait 16 time steps before starting to record the maxima, in order to wait for the system to fall in the first local minimum,
and $E_\MAX(t/2,t)$ is the maximum energy in the time window $[t/2,t]$. The latter has the advantage of showing very well the separation of time scales
that occurs during the dynamics, since it takes into account increasingly larger time windows and excludes configurations visited long ago.
In the run depicted, the initial state has $E(0)\approx0.2 N$, so $E_\MAX(0,t)$ sticks to that value for a long time. Instead,
$E_\MAX(16,t)$ starts from a negative value, since in 16 time steps the system had the time to explore a portion of the landscape,
lowering its energy. 
After around $10^{10}$ \ac{mc} steps $E_\MAX(16,t)$ reaches $E(0)$, and, from that moment on, $E_\MAX(16,t)=E_\MAX(0,t)$. This is what would be the time $t_3$, marking the beginning of the First Records regime.

$E_\MAX(t/2,t)$ is the only indicator that can also decrease. It is equal to $E_\MAX(16,t)$ either at the beginning of the run, when
the system is stuck in the initial trap, or when new records are hit. 
As stated previously, $E_\MAX(t/2,t)$ is a good separator of time scales. In fact, several times during the aging stage $E_\MAX(t/2,t)$ drops below $\inf I$, indicating that the dynamics has accepted no move and has remained in the current trap for the time interval considered.

\bibliographystyle{unsrt}
\bibliography{marco}

\end{document}

%% file: Emax_tw_tex.tex
\begingroup
  \makeatletter
  \providecommand\color[2][]{%
    \GenericError{(gnuplot) \space\space\space\@spaces}{%
      Package color not loaded in conjunction with
      terminal option `colourtext'%
    }{See the gnuplot documentation for explanation.%
    }{Either use 'blacktext' in gnuplot or load the package
      color.sty in LaTeX.}%
    \renewcommand\color[2][]{}%
  }%
  \providecommand\includegraphics[2][]{%
    \GenericError{(gnuplot) \space\space\space\@spaces}{%
      Package graphicx or graphics not loaded%
    }{See the gnuplot documentation for explanation.%
    }{The gnuplot epslatex terminal needs graphicx.sty or graphics.sty.}%
    \renewcommand\includegraphics[2][]{}%
  }%
  \providecommand\rotatebox[2]{#2}%
  \@ifundefined{ifGPcolor}{%
    \newif\ifGPcolor
    \GPcolorfalse
  }{}%
  \@ifundefined{ifGPblacktext}{%
    \newif\ifGPblacktext
    \GPblacktexttrue
  }{}%
  \let\gplgaddtomacro\g@addto@macro
  \gdef\gplbacktext{}%
  \gdef\gplfronttext{}%
  \makeatother
  \ifGPblacktext
    \def\colorrgb#1{}%
    \def\colorgray#1{}%
  \else
    \ifGPcolor
      \def\colorrgb#1{\color[rgb]{#1}}%
      \def\colorgray#1{\color[gray]{#1}}%
      \expandafter\def\csname LTw\endcsname{\color{white}}%
      \expandafter\def\csname LTb\endcsname{\color{black}}%
      \expandafter\def\csname LTa\endcsname{\color{black}}%
      \expandafter\def\csname LT0\endcsname{\color[rgb]{1,0,0}}%
      \expandafter\def\csname LT1\endcsname{\color[rgb]{0,1,0}}%
      \expandafter\def\csname LT2\endcsname{\color[rgb]{0,0,1}}%
      \expandafter\def\csname LT3\endcsname{\color[rgb]{1,0,1}}%
      \expandafter\def\csname LT4\endcsname{\color[rgb]{0,1,1}}%
      \expandafter\def\csname LT5\endcsname{\color[rgb]{1,1,0}}%
      \expandafter\def\csname LT6\endcsname{\color[rgb]{0,0,0}}%
      \expandafter\def\csname LT7\endcsname{\color[rgb]{1,0.3,0}}%
      \expandafter\def\csname LT8\endcsname{\color[rgb]{0.5,0.5,0.5}}%
    \else
      \def\colorrgb#1{\color{black}}%
      \def\colorgray#1{\color[gray]{#1}}%
      \expandafter\def\csname LTw\endcsname{\color{white}}%
      \expandafter\def\csname LTb\endcsname{\color{black}}%
      \expandafter\def\csname LTa\endcsname{\color{black}}%
      \expandafter\def\csname LT0\endcsname{\color{black}}%
      \expandafter\def\csname LT1\endcsname{\color{black}}%
      \expandafter\def\csname LT2\endcsname{\color{black}}%
      \expandafter\def\csname LT3\endcsname{\color{black}}%
      \expandafter\def\csname LT4\endcsname{\color{black}}%
      \expandafter\def\csname LT5\endcsname{\color{black}}%
      \expandafter\def\csname LT6\endcsname{\color{black}}%
      \expandafter\def\csname LT7\endcsname{\color{black}}%
      \expandafter\def\csname LT8\endcsname{\color{black}}%
    \fi
  \fi
    \setlength{\unitlength}{0.0500bp}%
    \ifx\gptboxheight\undefined%
      \newlength{\gptboxheight}%
      \newlength{\gptboxwidth}%
      \newsavebox{\gptboxtext}%
    \fi%
    \setlength{\fboxrule}{0.5pt}%
    \setlength{\fboxsep}{1pt}%
\begin{picture}(7200.00,5040.00)%
    \gplgaddtomacro\gplbacktext{%
      \csname LTb\endcsname%
      \put(946,1213){\makebox(0,0)[r]{\strut{}$-0.6$}}%
      \put(946,2231){\makebox(0,0)[r]{\strut{}$-0.4$}}%
      \put(946,3248){\makebox(0,0)[r]{\strut{}$-0.2$}}%
      \put(946,4266){\makebox(0,0)[r]{\strut{}$0$}}%
      \put(946,4775){\makebox(0,0)[r]{\strut{}$0.1$}}%
      \put(1078,484){\makebox(0,0){\strut{}$0$}}%
      \put(2223,484){\makebox(0,0){\strut{}$0.04$}}%
      \put(3368,484){\makebox(0,0){\strut{}$0.08$}}%
      \put(4513,484){\makebox(0,0){\strut{}$0.12$}}%
      \put(5658,484){\makebox(0,0){\strut{}$0.16$}}%
      \put(6803,484){\makebox(0,0){\strut{}$0.2$}}%
    }%
    \gplgaddtomacro\gplfronttext{%
      \csname LTb\endcsname%
      \put(176,2739){\rotatebox{-270}{\makebox(0,0){\strut{}\Large $E_{\text{max}}(t/2,t)/N$}}}%
      \put(3940,154){\makebox(0,0){\strut{}\Large $\log(t)/N$}}%
      \csname LTb\endcsname%
      \put(6428,3634){\makebox(0,0)[r]{\strut{}$N = 64$}}%
      \csname LTb\endcsname%
      \put(6428,3830){\makebox(0,0)[r]{\strut{}$N = 96$}}%
      \csname LTb\endcsname%
      \put(6428,4026){\makebox(0,0)[r]{\strut{}$N = 128$}}%
      \csname LTb\endcsname%
      \put(6428,4222){\makebox(0,0)[r]{\strut{}$N = 192$}}%
      \csname LTb\endcsname%
      \put(6428,4418){\makebox(0,0)[r]{\strut{}$N = 256$}}%
      \csname LTb\endcsname%
      \put(6428,4614){\makebox(0,0)[r]{\strut{}$N = 384$}}%
    }%
    \gplgaddtomacro\gplbacktext{%
      \csname LTb\endcsname%
      \put(2878,3363){\makebox(0,0)[r]{\strut{}$-0.72$}}%
      \put(2878,3778){\makebox(0,0)[r]{\strut{}$-0.66$}}%
      \put(2878,4193){\makebox(0,0)[r]{\strut{}$-0.6$}}%
      \put(3010,2852){\makebox(0,0){\strut{}$0$}}%
      \put(3763,2852){\makebox(0,0){\strut{}$0.04$}}%
      \put(4516,2852){\makebox(0,0){\strut{}$0.08$}}%
      \put(5269,2852){\makebox(0,0){\strut{}$0.12$}}%
    }%
    \gplgaddtomacro\gplfronttext{%
      \csname LTb\endcsname%
      \put(1976,3722){\rotatebox{-270}{\makebox(0,0){\strut{}$E_{\mbox{th}}/N$}}}%
      \put(4186,2522){\makebox(0,0){\strut{}$N^{-1/2}$}}%
      \csname LTb\endcsname%
    }%
    \gplbacktext
    \put(0,0){\includegraphics{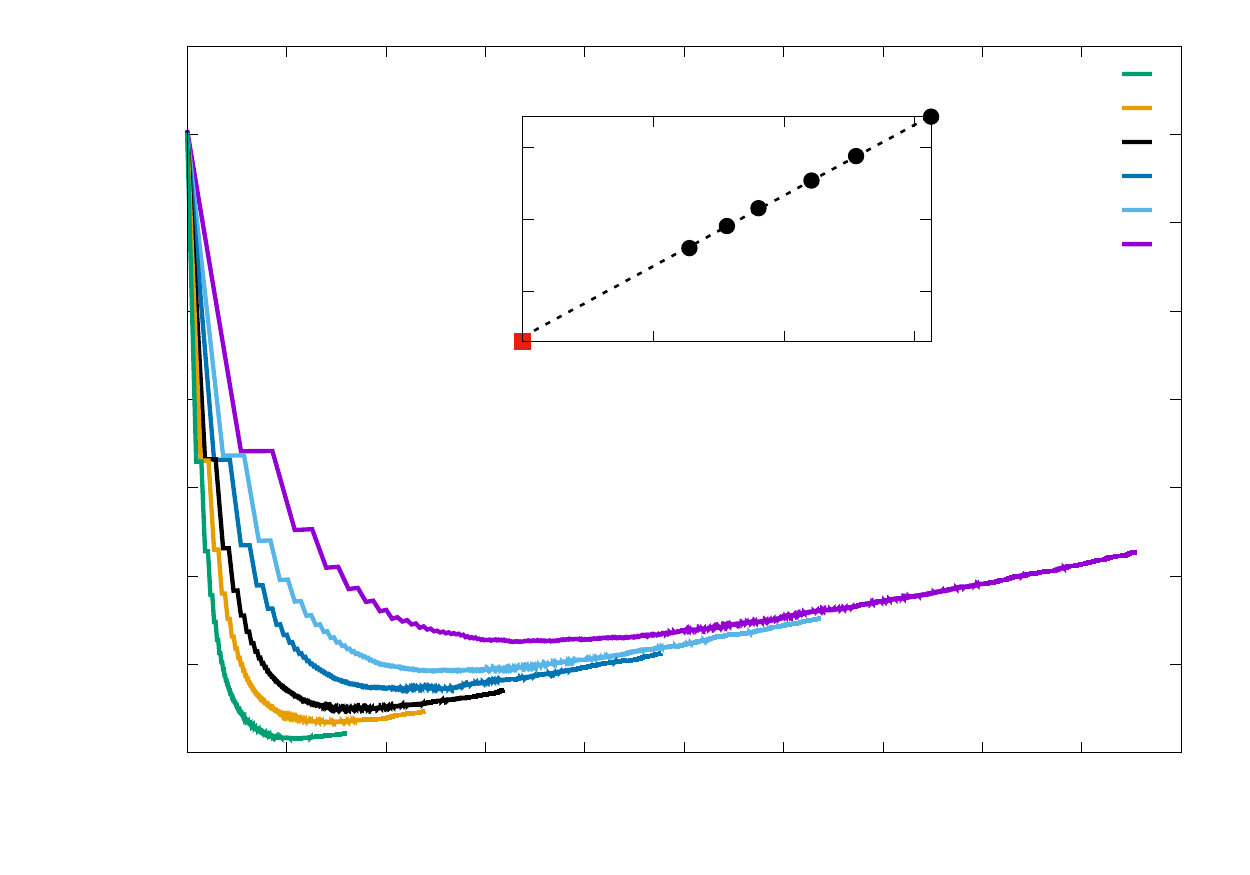}}%
    \gplfronttext
  \end{picture}%
\endgroup